\documentclass[12pt]{report}

\usepackage{amsmath,amssymb,amsfonts}
\usepackage[
  a4paper,
  top=1.25in, bottom=1.25in, outer = 1in,
  inner=1.2in
  ]{geometry}
\usepackage{multicol}
\usepackage{mdframed}
\usepackage{amsthm}
\usepackage{stmaryrd}
\usepackage{accents}
\usepackage{graphicx}
\usepackage{centernot}
\usepackage{tikz}
\usepackage{framed}
\usepackage{subcaption}
\usepackage{scalerel,stackengine}
\usepackage[hidelinks]{hyperref}
\usepackage{pict2e,picture}
\usepackage{mathtools}
\usepackage{proof}
\usepackage{enumerate}
\usepackage{cleveref}
\usepackage{subfiles} 

\usetikzlibrary{arrows.meta}

\newtheoremstyle{mytheoremstyle} 
    {\topsep}                    
    {\topsep}                    
    {\rm}                   
    {}                           
    {\bf}                   
    {.}                          
    {.5em}                       
    {}  

\theoremstyle{mytheoremstyle}

\newtheorem{theo}{Theorem}[chapter]
\newtheorem{prop}{Proposition}[chapter]
\newtheorem{deff}{Definition}[chapter]
\newtheorem{lemma}{Lemma}[chapter]
\newtheorem{example}{Example}[chapter]
\newtheorem{notation}{Notation}[chapter]
\newtheorem{remark}{Remark}[chapter]
\newtheorem{corollary}{Corollary}[chapter]

\DeclareFontFamily{U}{mathx}{\hyphenchar\font45}
\DeclareFontShape{U}{mathx}{m}{n}{
      <5> <6> <7> <8> <9> <10>
       <10.95> <12> <14.4> <17.28> <20.74> <24.88>
       mathx10
       }{}
 \DeclareSymbolFont{mathx}{U}{mathx}{m}{n}
 \DeclareFontSubstitution{U}{mathx}{m}{n}
 \DeclareMathAccent{\widecheck}{0}{mathx}{"71}
 \DeclareMathAccent{\wideparen}{0}{mathx}{"75}

\stackMath
\newcommand\reallywidecheck[1]{%
\savestack{\tmpbox}{\stretchto{%
  \scaleto{%
    \scalerel*[\widthof{\ensuremath{#1}}]{\kern.1pt\mathchar"0362\kern.1pt}%
    {\rule{0ex}{\textheight}}
  }{\textheight}%
}{2.4ex}}%
\stackon[2pt]{#1}{\scalebox{-1}{\tmpbox}}%
}
\stackMath
\newcommand\widetriangle[1]{%
\savestack{\tmpbox}{\stretchto{%
  \scaleto{%
    \scalerel*[\widthof{\ensuremath{#1}}]{\kern.1pt\triangle\kern.1pt}%
    {\rule{1ex}{\textheight}}
  }{\textheight}%
}{1ex}}%
\stackon[2pt]{#1}{\tmpbox}%
}
\newcommand{\ralg}{\mathcal A}
\newcommand{\ralgb}{\mathcal B}
\newcommand{\lattice}{\mathcal L}

\newcommand{\bottom}{\perp}
\newcommand{\halfvec}[1]{\accentset{\rightharpoonup}{#1}}
\newcommand{\rel}[3]{{\mathtt{#1}}\,\,{#2}\,\,{\mathtt{#3}}}

\newcommand{\op}{^\circ}

\newcommand{\antihat}{\widecheck}

\newcommand{\seqsym}{\mathsf{s}}
\newcommand{\parsym}{\mathsf{p}}
\newcommand{\fullsym}{\mathsf{h}}

\newcommand{\opsym}{\circ}

\newcommand{\seqclosure}[1]{#1^{\seqsym}}
\newcommand{\parclosure}[1]{#1^{\parsym}}
\newcommand{\fullclosure}[1]{#1^{\fullsym}}

\newcommand{\dotleq}[1]{\,\,\dot\leq_#1\,\,} 

\newcommand{\dotleqx}{\,\,\dot\leq\,\,} 

\newcommand{\longantihat}{\reallywidecheck}

\newcommand{\deriv}[2]{#1\langle #2 \rangle}

\newcommand{\compwise}[1]{\boldsymbol{#1}}

\newcommand{\termset}{\mathcal{S}V}

\newcommand{\tay}[2]{\partial^{#1}(#2)}

\newcommand{\taynpar}[2]{\partial^{#1}#2}

\newcommand{\myinfer}[2]{\infer {#2} {#1}}

\newcommand{\subst}[2]{{#2}^{#1}}

\newcommand{\logicand}{\,\,\&\,\,}

\newcommand{\definedas}{\overset{\text{\tiny$\triangle$}}=}
\newcommand{\trianglearrow}{
\begin{tikzpicture}[outer sep=0, inner sep = 0,baseline={([yshift=-.6ex]current bounding box.center)}]
 \node (a) at (0,0) {};
 \node (b) at (.5, 0){};
 \draw [-{Triangle[open]}] (a) -- (b);
\end{tikzpicture}
}

\newcommand{\repltbar}{x}
\newcommand{\replsbar}{y}
\newcommand{\replrbar}{z}

\newcommand{\antihatadj}[1]{\Diamond(#1)}

\newcommand{\so}[1]{#1^{\seqsym\opsym}}
\newcommand{\os}[1]{#1^{\opsym\seqsym}}

\newcommand{\ooso}[1]{#1^{\opsym\opsym\seqsym\opsym}}
\newcommand{\osoo}[1]{#1^{\opsym\seqsym\opsym\opsym}}
\newcommand{\soo}[1]{#1^{\seqsym\opsym\opsym}}
\newcommand{\derivdelta}{\deriv \Delta}

\newcommand{\Rel}[1]{Rel_{#1}}

\newcommand{\relone}{a}

\newcommand{\bolditem}[1]{\item \textbf{#1}\quad}

\newcommand{\idrel}{\Delta}

\title{A Relation Algebra for Term Rewriting \\
\Large{A differential approach to sequential reduction}\\
\Large (Revised Version)}
\author{Lorenzo Pace}

\begin{document}
\pagenumbering{roman}
\maketitle


\section*{Foreward}
This is a revised version of the author's BSc dissertation, written under the supervision of Filippo Bonchi and Francesco Gavazzo 
and defended in October 2023 at the University of Pisa.

\begin{abstract}
    Recently, Gavazzo has developed a \emph{relational theory of symbolic manipulation}, that allows to study syntax-based rewriting systems without relying on specific notions of syntax. This theory was obtained by extending the algebra of relations with syntax-inspired operators. Within the algebras thus obtained, it is possible to encode notions of parallel and full reduction for first-order rewriting systems, 
as well as to prove nontrivial properties about them in an algebraic and syntax-independent fashion.
     \emph{Sequential} reduction, however, was not explored, but it was conjectured that it could be studied through a \emph{differential} relational theory of rewriting. This manuscript proves the above conjecture by defining \emph{differential algebras of 
    term relations}, viz. algebras of term relations 
    extended with novel operators inspired by the theory of functor derivatives. 
    We give a set of axioms and rules for such operators and show that the resulting theory is expressive enough to 
    define notions of parallel, full, and \emph{sequential} reduction.  
    We prove fundamental results relating all these notions in a purely algebraic and syntax-independent way, 
    and showcase the effectiveness of our theory by proving the soundness of a proof technique for weak confluence 
    akin to the so-called Critical Pair Lemma.
\end{abstract}

\clearpage
\tableofcontents

\clearpage

\pagenumbering{arabic}








\chapter{Introduction}

\label{chapter:intro}
\section{Motivation and Outline}
 \paragraph{Rewriting theory} Rewriting theory \cite{newman, terese} studies \emph{symbolic manipulation}, that is, given a set of objects and a set of rules to transform one object into another, it studies the properties of these transformations. If the objects are formulas of any sort, this theory lets us reason about equational proofs, finding applications in fields such as automated and computer-aided theorem proving  \cite{HSIANG199271}. If they are programs, or abstract syntax trees, it lets us reason about program equivalence, and it finds several applications, for example:
 \begin{itemize}
  \item Compilers are able to rearrange a program's instructions to perform optimizations. For example, if a C compiler recognizes that an idempotent operation is placed within a \texttt{while} block, it may generate code that always only performs such operation once, by bringing the operation outside of the block. This technique is known as \emph{loop-invariant code motion} or \emph{hoisting} \cite{alfred2007compilers}.
  \item Given a mathematical computation, writing the same formula in two different yet equivalent ways may impact the precision significantly, as some machine operations are less stable than others \cite{higham2002accuracy}. Rewriting programs can thus not only improve the performance, but also the precision of our computations.
 \end{itemize}
Among the desireable properties of a rewriting system are \emph{confluence} \cite{Huet80}, that certifies the semi-decidability of the problem of asking whether two objects are equivalent\footnote{This sort of problem is often referred to as a \emph{word problem}.}, and \emph{termination} (also known as \emph{strong normalization} and \emph{Noetherianness})\cite{Huet80}, that assures us that every object can reach some irreducible ``normal form'' by repeated application of rules (i.e. that there are no \emph{loops} or infinite chains of rule applications), and together with confluence determines the decidability of the equivalence problem.

 \paragraph{Term Rewriting} Rewriting theory comes in many flavours, that can be partitioned into \emph{abstract} and \emph{syntax-based} rewriting. The former can be studied by itself, without any need for syntax, while that is not the case for the latter. 
 Any rewriting system comprises of \emph{objects} and \emph{rewrite rules}: when the objects are first-order terms, and the rewrite rules consist of a \emph{ground reduction} relation between terms, we obtain (first-order) \emph{Term Rewriting} \cite{terese}. Traditional texts on term rewriting usually introduce several syntax-based notions, that are not relative to rewriting in themselves, but are preliminary to the introduction of term-specific properties. Among these are:
 \begin{itemize}
  \item \emph{Occourrences} of the same sub-term within a term, e.g.:
  \[ \underbrace{(x + y)}_{\#1} \cdot 7 \cdot \underbrace{(x + y)}_{\#2} \]
  \item \emph{Contexts}, i.e. terms with a ``hole'', e.g.:
  \[ {(x + y)} \cdot (\square + 1) \]
  \item \emph{Substitution} of variables within terms, e.g. assume that $\sigma$ substitutes the variable $x$ with the expression $(10 \cdot y)$:
  \[ \sigma\big(( x + 5) \cdot x \big) =  (\underline{(10 \cdot y)} + 5) \cdot \underline{(10 \cdot y)} \]
 \end{itemize}

 \paragraph{The problem with Term Rewriting Theory} The limitation of first-order term rewriting theory lies in its difficulty to generalize certain term-specific notions to other related rewriting systems, such as higher order term rewriting systems (e.g. the $\lambda$-calculus) and graph rewriting systems. The reason behind this flaw is the syntax-dependance of these notions (particularly evident in the notion of \emph{occourrence}). One notable example of such a tedious-to-generalize result is the \emph{critical pair lemma} \cite{Huet80}, which states that a term rewriting system is confluent if and only if every critical pair\footnote{The definition of a critical pair (CP) is quite convoluted, and requires notation that has not yet been introduced. Intuitively, one can think of a CP as a pair of terms that can be obtained by applying two different rules to a single other term. The actual definition is more nuanced than this, e.g. it excludes pairs that can be obtained by substitution from other CPs.} converges. Each time that an analogous lemma is to be proven for other rewriting systems, it is necessary to start from scratch.

%

\paragraph{Can we do better?} In order to study these concepts in a way that is easier to generalize, we need to employ a more abstract theory of rewriting. This new theory of rewriting should still be able to express all term-specific notions, but without any mention of syntax.

\begin{itemize}
 \item The first candidate is the theory of \emph{Abstract Rewriting Systems} \cite{terese}, a more general theory of rewriting that allows us to capture several general rewriting notions, such as confluence and termination. Remarkably, it has been shown that abstract rewriting can be studied algebraically through Algebras of Relations \cite{tarski-1941}. However, this theory is not expressive enough to obtain term-specific results such as the critical pair lemma.

 \item Our approach builds on the algebraic, relational approach to rewriting mentioned earlier, but extends it with operators that model the term-specific notions in a completely syntax-independent fashion.
\end{itemize}

%

%

\paragraph{Algebras of Term-Relations}
Recently Gavazzo \cite{Gavazzo/LICS/2023} has extended the algebra of relations with syntax-inspired operators that model term-specific notions. The strength of the purely algebraic approach is that it does not only apply to reductions between terms, but can be generalized to analogous structures that can be studied through the tools of Category Theory. Following this approach, it was possible to obtain an \emph{Algebra of Term-Relations} that models two specific kinds of reductions, namely \emph{parallel} and \emph{full} (or \emph{deep}) reduction. The former allows for parallel application of independent rewrite rules, while the latter allows for parallel application of non-necessarily-independent (i.e. possibly nested) rewrite rules. Of course, this theory is missing a piece; it does not model the sort of reduction that performs only one step at a time, to which we refer as \emph{sequential reduction}. 

\paragraph{Original Contribution} 

In section X of \cite{Gavazzo/LICS/2023} (\emph{Sequential reduction: A Few Words Only}), the author mentions that:

\begin{quote}
[...] The theory developed so far shows that parallel and full
reduction are remarkably natural, at least from a structural and
algebraic perspective. When it comes to think about reduction
computationally, however, \emph{sequential} (or \emph{linear}) reduction
is usually considered more fundamental. [...]
\end{quote}\noindent
and later goes on to conjecture that it could be possible to study sequential reductions through operators inspired by \emph{differential calculus}.

This manuscript aims to refine Gavazzo's Algebra of Term-Relations by introducing additional operators to model sequential reduction and integrating them with previously defined operators to provide a comprehensive spectrum of reductions. Additionally, we prove the above conjecture correct, as discussed in Section \ref{section:mathematical_analysis_of_syntax}. The operators we define are inspired by categorical notions, yet they can also be obtained purely algebraically.

\paragraph{List of contributions} The main contributions of this manuscript follow:
\begin{itemize}
    \item The definition of sequential refinement and sequential closure.
    \item Their algebraic formalization through \emph{Sequential Algebras of Term-Relations};
    \item The concrete definitions of derivative and Taylor expansion of a term-relation;
    \item Their algebraic formalization through \emph{Differential Algebras of Term-Relations};
    \item The proof of two fundamental theorems that link sequential and parallel reduction;
    \item The application of the newly defined algebra to the proof of a Critical Pair-like lemma.
\end{itemize}

\paragraph{Outline of the document}
\begin{itemize}
 \bolditem {Chapter \ref{chapter:mathprel}}
 In this chapter we recall some mathematical preliminaries, including lattices, adjunctions, the \emph{fixed-point calculus} \cite{mathematics1995fixed}, and relations. These are all the tools we shall employ in the formal proofs throughout the document.
\bolditem {Chapter \ref{chapter:abstract_rewriting}} This chapter introduces abstract rewriting, first from a traditional perspective and then through a relational lens. We illustrate the expressivity of this approach by providing a relational proof of a classic result: the equivalence between confluence and the Church-Rosser property.

\bolditem {Chapter \ref{chapter:term_rewriting}} In this chapter we present Term Rewriting Systems initially using the traditional viewpoint. Later, we redefine them in the context of $\ralg$-TRSs, applying Gavazzo's theory of Algebras of Term Relations \cite{Gavazzo/LICS/2023}.

\bolditem {Chapter \ref{chapter:seq_red}} The original contribution of this work begins in this chapter, where we introduce the \emph{sequential refinement} operator concretely (i.e. still relying on syntax to some extent) and develop the theory of \emph{Sequential Algebras of Term-Relations}, allowing for a purely algebraic approach to sequential reduction. 

\bolditem {Chapter \ref{chapter:DAoR}} In this chapter we initially introduce the \emph{Derivative} and \emph{Taylor Expansion} operators concretely. Subsequently, we establish the broader theory of \emph{Differential Algebras of Term-Relations} and prove two \emph{fundamental theorems} that bridge the theory of sequential reduction with that of parallel reduction. The final result of this chapter shows a \emph{spectrum} of reductions, illustrating how, for example, parallel and full reductions can be simulated through a finite amount of sequential steps. The reason behind the surprising borrowing of nomenclature from the field of Mathematical Analysis is justified in Section \ref{section:mathematical_analysis_of_syntax}.

\bolditem {Chapter \ref{chapter:CP-like}} In the final chapter, we put our newly defined theory into practice, culminating in the proof of a result akin to the Critical Pair lemma. This way we showcase the elegance and generality of the algebraic approach to rewriting.
\end{itemize}

{ 
\newcommand{\signature}{\Sigma}
\newcommand{\variables}{V}
\newcommand{\termone}{\mathtt{t}}
\newcommand{\defeq}{\definedas}
\newcommand{\bbm}{\mathbb} 
\newcommand{\termtwo}{{\tt s}}
\newcommand{\catname}[1]{\textsf{#1}}

\renewcommand{\op}{o}
\newcommand{\constant}{\mathtt{k}}
\newcommand{\varone}{\mathtt{x}}
\newcommand{\quot}[2]{\left.\raisebox{.2em}{$#1$}\middle/\raisebox{-.2em}{$#2$}\right.}

\section{From Derivatives to Algebra: the differential nature of sequential reduction}
\label{section:mathematical_analysis_of_syntax}

Before moving to the technical development of our theory, it is instructive to 
explain why we refer to our approach as \emph{differential}. Unfortunately, a precise explanation 
requires a moderate acquaintance with basic category theory~\cite{MacLane/Book/1971} 
as well as the introduction of several non-basic categorical concepts that significantly go beyond 
the scope of this manuscript. Consequently, we shall try to give the reader an informal idea of 
the link between sequential reduction and (functor) derivatives, in a general sense. Doing so is not easy, 
and oftentimes we will develop our arguments using basic term rewriting formalism, although we will 
try to do so in an intuitive and minimal way. The reader is invited to read this section without trying to 
understand all the details, and then come back to it after having read the core parts of the work: hopefully, 
this process will shed some light on the link between sequential reduction and (functor) derivatives.


\subsection{From Categories to Allegories, to Algebra}
Let us begin with an explanation of the shift from category theory to `traditional' algebra, the latter being 
the formal framework employed in this manuscript.
Gavazzo's approach to relational rewriting~\cite{Gavazzo/LICS/2023} begins with the categorical analysis of 
syntax~\cite{Goguen/1977,fiore-plotkin-turi-99}, whereby the abstract syntax of a language is modeled 
relying on categorical notions such as (polynomial) functors, monads, and initial algebras. 
Gavazzo observed that the very same categorical notions used to model syntax are exactly what is needed 
to model notions of reductions between syntactic expressions, provided one transfers such 
notions to a \emph{relational} base. In set-theoretic terms -- the only needed for this manuscript -- 
that means that one needs to extend the action of 
notions acting on functions to \emph{relations}. 

\paragraph{An Informal Example}
To clarify the concept, it is useful to introduce a simple example (the reader not familiar with the notation 
is invited to skip this section and come back to it after having read Chapter \ref{chapter:term_rewriting}). 
Let us consider a rudimentary language for arithmetic made of three operations: a constant 
${\tt 0}$ for zero, a unary operation ${\tt S}$ for the successor function, and a binary operation 
${\tt A}$ for addition. These data are what constitute the \emph{signature} of the language. 
In fact, once we fix a set $V$ of variables, we have all we need to build expressions: starting from 
variables, we combine expressions using operations, this way building increasingly complex expressions. 

For instance, if $x$ is a variable, then $x$, ${\tt 0}$, ${\tt S}(x)$, ${\tt A}({\tt 0}, {\tt S}(x)), \hdots$ 
are all expressions in the language of arithmetic. 
All of that can be made mathematically rigorous by employing simple categorical notions. For instance, 
we can view signatures as functors. In our case, we have a construction 
$\signature$ actions on sets defined thus:
\[\signature(A) \defeq \{  {\tt 0}, {\tt S} (\termone_1), {\tt A} (\termone_1, \termone_2)  \mid \termone_i \in A \}\]
Being a functor, $\signature$ also acts on functions: whenever we have a function $f: A \to B$, we also have a 
function $\signature(f): \signature(A) \to \signature(B)$ obtained by applying $f$ along operators 
(e.g. $\signature(f)({\tt A}(\termone_, \termone_2)) = {\tt A}(f(\termone_1), f(\termone_2))$).
Even if this is not precise, it is evident that by using such notions, it is possible to 
recover many syntactic notions in a mathematically precise way. As an example, once the signature functor is given, 
we can consider its nested applications $\signature(\signature(\cdots \signature(A) \cdots))$, 
so that by taking 
$\bigcup_{n \geq 0} \signature^n(V)$ we recover the set $T_\signature (V)$ of expressions in the language of arithmetic. 

\paragraph{Relational Rewriting} 
Now, what does all of that have to do with rewriting? The starting point of relational rewriting 
is the observation that the aforementioned set-and-function notions used to `mathematize' the theory 
of syntax, can be used to mathematize the theory of syntactic transformations (viz. reductions) if extended 
to set-and-\emph{relation}. That means, in particular, that not only the set constructor $\signature$ has 
to act on functions, but also on (binary) relations: whenever we have a relation 
$\relone \subseteq A \times B$, we must also have a relation 
$\signature(\relone) \subseteq \signature(A) \times \signature(B)$. 
Continuing our running example, the definition of $\signature(\relone)$ is quite natural: 
two expressions are related by $\signature(\relone)$ if they have the same outermost operation and pairwise 
$\relone$-related arguments. At this point, the reader may have an intuition of what is next: if we let $\relone$ model 
a notion of reduction on $A$, then $\signature(\relone)$ somehow gives a notion of parallel reduction on 
`expressions' over $A$. 

Remarkably, there is a rich theory of extensions of set-functors to 
relations~\cite{Kurz/Tutorial-relation-lifting/2016} that has also been generalized 
to arbitrary categories and \emph{allegories}~\cite{scedrov-freyd} -- the latter being 
a relational counterpart of the notion of a category --
in terms of \emph{relational extensions}~\cite{Barr/LMM/1970,carboni-kelly-wood} and 
\emph{relators}~\cite{kawahara1973notes,backhouse-polynomial-relators,algebra-of-programming}. 
Building on top of such a theory, Gavazzo has extended categorical theories of syntax to 
allegories, showing how the resulting relational/allegorical notions are precisely the building block 
needed to define general rewriting theories on syntax-based systems. 

In doing so, Gavazzo realized that the categorical and allegorical machinery is 
actually not needed, as one could take the aforementioned relational notions and their 
structural properties as primitive operators and axioms, respectively, in a traditional algebra of 
relations in the spirit of Tarski~\cite{tarski-1941}. 
The result of that is an extended algebra of relations, called \emph{algebra of term} (or \emph{program}) \emph{relations}, 
that makes no reference to categories, allegories, etc. and inside of which a rich theory of 
(syntax-based) rewriting can be developed in an algebraic and calculational style, and, most importantly, 
 in a completely \emph{syntax-independent} fashion. 
 The categorical-allegorical construction used to identify the operators and axioms of the algebra of term 
 relations can be then seen as a general model existence theorem that, once instantiated on 
 the category of sets and functions, and the allegory of sets and relations, precisely gives 
 the traditional theory of term rewriting systems.

\paragraph{From Functors to their Derivatives}
Algebras of term relations come with operators that are expressive enough to fruitfully model notions of 
\emph{parallel reduction}. This comes with no surprise if we think about the previously given 
definition of $\signature(\relone)$. In more general terms, 
relational extensions and relators being relational counterparts of functors are inherently parallel,\footnote{
Thinking about elements of $F(A)$ for a functor $F$ as kind of $F$-trees over $A$, we see that 
$F(f)(t)$ applies $f$ to all leaves of $t$ in a parallel fashion.} 
and thus it is unclear how to model sequential notions of reduction. Of course one possibility is to proceed 
from scratch ignoring relational extensions, but that forces one to throw away a rich and powerful body 
of theories and results from which a relational theory of rewriting would benefit. 

As already mentioned, in \cite{Gavazzo/LICS/2023} it is conjectured that 
a theory of sequential reduction can be still developed in terms of allegories, 
relational extensions, and, ultimately, algebras of term relations: to do that, however, 
we do not have to work with signature functors and related notions, but rather with 
their \emph{derivatives}.

\subsection{Enter the Derivative}
Let us look again at our running example. Suppose to have a set $A$ with a notion of reduction $\mapsto$ on it, i.e. 
a relation ${\mapsto} \subseteq A \times A$. We have already seen that 
$\signature(\mapsto)$ then gives 
a reduction on $\signature(A)$: accordingly, $\signature(\mapsto)$ reduces expressions in parallel, so that 
if $\termone_1 \mapsto \termtwo_1$ and $\termone_2 \mapsto \termtwo_2$, then:
\[{\tt A}(\termone_1, \termone_2) \mathrel{\signature(\mapsto)} {\tt A}(\termtwo_1, \termtwo_2).\] 
To reduce ${\tt A}(\termone_1, \termone_2)$ sequentially, we should fix a specific position 
indicating where reduction should happen. This could be done by employing a special symbol 
$\Box$ (a \emph{hole}) and by decomposing a term $\termone$ as pair consisting of term 
${\tt C}$ with \emph{exactly one} occurrence of $\Box$ and a term $\termone_0$ such that 
replacing $\Box$ with $\termone$ in ${\tt C}$ gives us exactly $\termone$. 
This way, ${\tt C}$ tells us where the sequential reduction happens, whereas $\termone_0$ 
gives the term that has to be reduced. 
Expressions such as ${\tt C}$ are known as \emph{linear contexts} and they can be thought of as 
kind of \emph{derivatives} of expressions. 

\paragraph{Sequentiality} 
McBride \cite{derivative-1} observed that to any signature (functor) $\signature$, we can associate 
another set-based (functorial) construction $\partial \signature$ that maps any set $A$ to 
the collection of \emph{linear contexts} built out of $\signature A$. 
In our language for arithmetic, examples of elements of $\partial\signature(A)$ are: 
${\tt S}(\Box)$, ${\tt A}(\Box, {\tt S}({\tt Z}))$, and 
${\tt A}({\tt S}({\tt S}({\tt 0})), ({\tt S}({\tt \Box}))$. 
Moreover, there is a ``plugging-in" map: \[@: \partial\signature(A) \times A \to \signature(A)\] that maps a context ${\tt C}$ and an element ${\tt t}$ in $A$ to the element of $\signature(A)$ 
obtained by replacing the hole in ${\tt C}$ with ${\tt t}$. 

At this point, it should come as no surprise that sequential reduction can be obtained 
in terms of relational extensions of the construction 
$\partial \signature(-) \times -$. Given a relation $\relone$, the relation 
$\partial\signature(\relone)$ behaves exactly as $\signature(\relone)$ except that 
it leaves the hole $\Box$ unchanged (thus, for instance, if $\termone \mapsto \termtwo$, then 
${\tt A}(\termone, \Box) \mathrel{\partial\signature(\mapsto)} {\tt A}(\termtwo, \Box)$). 
Consequently, $\partial \signature(\relone) \times \relone$ relate two pairs 
$({\tt C}_1, \termone_1)$ and $({\tt C}_2, \termone_2)$ if 
$\partial\signature(\relone)$ relates ${\tt C}_1$ with ${\tt C}_2$ and 
$\relone$ relates $\termone_1$ with $\termone_2$.

Putting the pieces together, we start with a reduction 
$\mapsto$, and to sequentially reduce an expression $\termone \in A$ we look at its (possibly many) decompositions 
as an element in $\partial \signature(A) \times A$, apply $\partial \signature(\mapsto) \times {\mapsto}$,
and then perform the plug-in operation. 
Thinking more carefully about the previous sentence, we realise that 
$\partial \signature(\mapsto) \times {\mapsto}$ actually reduces both the linear context and its argument, hence 
giving back a kind of parallel reduction. Fixing the issue is straightforward; we just consider 
$\partial \signature(\idrel) \times {\mapsto}$, where $\idrel$ is the identity relation. 
More generally, what we obtain is a binary operation on relations that given relations $\relone$ and 
$b$, returns $\partial \signature(\relone) \times {b}$. Sequential reduction is ultimately 
obtained by fixing the first argument to the identity relation but, from a meta-theoretical perspective, 
we will see that working with a binary operator is crucial. For instance, the reader should quite persuaded 
that one can recover $\signature(\relone)$ (almost) as $\partial \signature(\relone) \times {\relone}$. 

\paragraph{Derivatives and Linearity} 
At this point, the reader may rightfully ask: \emph{where are the derivatives?} 
The answer to this question is twofold: on the one hand, as observed in 
\cite{derivative-1}, the construction $\partial\signature$ obeys the laws of 
differential calculus (viz. chain rule, rule of product, etc.), which already suggests that $\partial\signature$ behaves like a sort of derivative of 
$\signature$.
On the other hand, and most importantly for us, it is possible to take an operational view on 
derivatives~\cite{spivak1971calculus} and think about the derivative of a 
 function $f$ as the best local \emph{linear approximation} of $f$.  
 Here, linearity is the key.

 Starting with the seminal work by Girard~\cite{DBLP:journals/tcs/Girard87}, 
researchers have recognized similarities 
between linearity as commonly used in, 
e.g. algebra and calculus, and linearity in the sense of logic, whereby a function is linear if it uses its arguments exactly once. 
In recent years, connections between logical and `analytical' linearity have been pushed much further, with the recent discovery by 
Erhahrd \cite{ehrhard2003differential}
that differentiation and linear function application, together with the associated notion of linear substitution, 
are deeply connected; in a suitable structural sense, they are the same. 
The aforementioned plug-in map $@: \partial\signature(A) \times A \to \signature(A)$ 
is linear in a suitable categorical sense\footnote{It is cartesian~\cite{derivative-2}.}; moreover, 
it gives a form of linear substitution (of an expression for the hole): 
all of that highlights a connection between the construction 
$\partial \signature$ (and $\signature$) 
and the concept of a derivative.\footnote{A precise account of such a connection 
was first discovered by Joyal~\cite{derivative-joyal}
in his theory of combinatorial species, and 
subsequently rediscovered by McBride and collaborators \cite{derivative-1,derivative-2} 
who gave a computational interpretation of the work by Joyal by thinking about \emph{derivatives as one hole contexts}.}

\paragraph{Taylor Expansion} 
The differential perspective on sequential reduction becomes even stronger if we apply 
the relational methodology not just on derivatives of signature functors, but also on their 
\emph{Taylor expansion} \cite{gylterud2011symmetric}.

In terms of syntax, the Taylor expansion of a signature functor expresses the idea that any expression can be seen 
an an $n$-ary context (i.e. a context with exactly $n$ holes), where $n$ can possibly be zero, with $n$ expressions 
plugged-in the context.
To be more precise, we have defined the derivative of a signature as a means to obtain the set of all one-hole contexts. By differentiating further we obtain the \emph{second derivative}, which naturally gives us the set of all two-hole contexts. Similarly, we can define the 
third derivative, and so on and so forth. For instance, writing $\partial^n\signature$ for the $n$-th derivative of 
$\signature$, our running example gives:
$$
\partial^2\signature(A) = \{{\tt A}(\Box_1, \Box_2), {\tt A}(\Box_2, \Box_1)\}.
$$
Can we use $\partial^n\signature(A)$ to perform the aforementioned decomposition of elements of 
$\signature A$? Not really, as having permutations of the placeholders is quite redundant: 
for instance, plugging-in $\termone_1,\termone_2$ in ${\tt A}(\Box_1, \Box_2)$ is the same as 
plugging-in $\termone_2, \termone_1$ in ${\tt A}(\Box_2, \Box_1)$. Stated otherwise, 
we can always recover the action of ${\tt A}(\Box_2, \Box_1)$ on a pair $(\termone, \termtwo)$ 
simply by given the reversed pair $(\termtwo, \termone)$ to ${\tt A}(\Box_1, \Box_2)$. 
This shows that there are redundancies in $\partial^n\signature(A)$.
To remove such redundancies it is enough to remove permutations of holes, hence quotienting 
$\partial^n\signature(A)$ with respect to the equivalence relation $\sim$ that 
relates couples of $n$-hole contexts if and only if they are the same context modulo permutations of the placeholders.

We shall denote the set $\quot{\partial^n\signature(A)}{\sim}$ as $\quot{\partial^n\signature(A)}{!n}$, due to 
the link of $\sim$ with the group of permutations on $n$-hole contexts (the cardinality of such a group being 
exactly $n!$). This way, we recover the desired decomposition as:
\[ \Sigma A \cong \coprod_{n \geq 0} \Bigg(  \dfrac{(\partial \Sigma)^n(\emptyset)}{!n} \times A^n \Bigg)\]
Note that the above equivalence uncannily resembles the definition of the Taylor-Maclaurin Series from Real Analysis \cite{trench2013introduction}:
\[ f(x) = \sum_{n = 0}^{\infty}\bigg( \dfrac{f^{(n)} (0)}{n!} \cdot x^n \bigg).\]

Similarly to what we did for derivatives, we can now look at the relational extension of 
$\coprod_{n \geq 0} \Bigg(  \dfrac{(\partial \Sigma)^n(\emptyset)}{!n} \times -^n \Bigg)$ and stratify 
using the relational extensions of 
$ \dfrac{(\partial \Sigma)^n(\emptyset)}{!n} \times -^n$, for each $n$.
Remarkably, such extensions define a form of parallel reduction where the reduction is applied in parallel exactly 
$n$-times; the Taylor expansion equivalence then tells us that a parallel reduction is exactly an $n$-parallel reduction, for 
some $n \geq 0$. This result, whose informal interpretation is astonishingly simple, gives an elegant and powerful bridge 
connecting sequential and parallel reduction.

\subsection{Back to Algebra}
All the categorical-oriented machinery informally discussed shows how the relational operators 
necessary to define sequential rewriting pop up, this way confirming Gavazzo's conjecture that 
sequential reduction can be recovered via (relational extensions of) derivates of functors. 
However, when it comes to actually studying rewriting systems it is more convenient to leave categorical 
and allegorical considerations under the hood, and to work axiomatically following the algebra of 
term relations methodology. This is precisely what we will do in this manuscript. 
In fact, once expressive enough algebras of term relations have been identified, 
one can completely forget about categorical and allegorical 
structure and (fruitfully) work in the framework of traditional algebra. 
What we shall do, then, is to algebraically axiomatize in the framework of algebras of term relations 
all the operators needed to reason about sequential reduction. Such operators and their axioms 
indeed ultimately correspond to the
relational extensions and their structural properties seen in this section, respectively. However, working 
algebraically we do not have to deal with the categorical bureaucracy, this way having shorter definitions and more 
effective proofs. Nonetheless, we will still do some work to justify our axioms and definitions; but instead of 
doing so in terms of categories, functors, etc., we look at the concrete example of term rewriting systems.

}



\chapter{Mathematical Preliminaries}
\label{chapter:mathprel}
\section{Lattices}

This section briefly introduces the fundamental concepts of lattice theory, which shall be used extensively throughout the following chapters. For a more complete account on the topic, see \cite{DaveyPriestley/Book/1990}.

\begin{deff} Given a partially ordered set $(P, \leq)$:
 \begin{itemize}
  \item The \emph{supremum} (or \emph{join}) of a subset $S$ of $P$ is the least (w.r.t. $\leq$) element $x$ such that $s \leq x$ for all $s \in S$. Such an element, if it exists, is denoted as $\bigvee S$.

 \item The \emph{infimum} (or \emph{meet}) of a subset $S$ of $P$ is the greatest (w.r.t. $\leq$) element $x$ such that $x \leq s$ for all $s \in S$. Such an element, if it exists, is denoted as $\bigwedge S$.

 \item $(P, \leq)$ is a \emph{lattice} iff any two-element subset $\{a, b\} \subseteq P$ has a join $a \vee b$ and a meet $a \wedge b$.

\item If $P$ is a lattice, it is said to be \emph{complete} when $\bigvee S$ and $\bigwedge S$ exist for any subset $S$ of $P$.
 
\end{itemize}
\end{deff}

\begin{example}
In the partial order $(\Rel A, \leq)$ of binary endorelations over a set $A$, where $\leq$ is the subset inclusion $\subseteq$, the join and meet of a set of relations $\{r_i \mid i \in I\}$ are their union and intersection respectively:
\[ \bigvee \{r_i \mid i \in I\} = \bigcup \{r_i \mid i \in I\}
\qquad \text{ and } \qquad \bigwedge \{r_i \mid i \in I\} = \bigcap \{r_i \mid i \in I\}.\]
%
%
%
%
\end{example}

\begin{deff} Given a partially ordered set $(P, \leq)$, we say that $y \in P$ is a \emph{top} (resp. \emph{bottom}) element of $P$ if $x \leq y$ (resp $y \leq x$) for any $x \in P$. Top and bottom elements, if they exist, are unique, and thus we denote them by $\top$ and $\bot$ respectively.
\end{deff}

\begin{example}
 In $(\Rel A, \leq)$, the top element is the relation $\nabla = \{(x, y) \mid x, y \in A\}$, whereas the bottom element is $\emptyset$.
\end{example}

Whenever we approach an algebraic structure, it is essential to think about which functions we should consider \emph{well-behaved}. Usually, the features of a well-behaved function have to do with structure preservation. 
                

  \begin{deff} 
  Let $(\lattice_1, \leq_1, \vee, \wedge)$, $(\lattice_2, \leq_2, \sqcup, \sqcap)$ be lattices and $F:\lattice_1 \to \lattice_2$ be a function.
  \begin{enumerate}
   \item $F$ is said to be \emph{monotonic} if:
  \[a \leq_1 b \implies F(a) \leq_2 F(b)\]
  i.e. it preserves the ordering.
  \item $F$ is said to be \emph{continuous} if:
  \[ F \left(\bigvee_i a_i\right) = \bigsqcup_i F(a_i) \]
  i.e. if it preserves joins.
  \end{enumerate}
  \end{deff}

%
%
%

 \begin{remark}
  \label{lax_pres_join}\emph{Lax} preservation of joins is equivalent to monotonicity:
 \[ \bigsqcup_i F(a_i)  \leq_2 F \left(\bigvee_i a_i\right) \iff \text{ $F$ monotonic.} \]
%
%
 Therefore, for a monotonic function, continuity is equivalent to \emph{oplax} preservation of joins.
 \end{remark}

\paragraph{$\omega$-chains}
It is often convenient to relax the notion of continuous function  to the one of $\omega$-continuous function.

\begin{deff}[$\omega$-chain]
An $\omega$-chain is a sequence $(a_n)_{n \geq 0}$ of elements of a lattice $\lattice$ such that, for all $i \geq 0$, $a_i \leq a_{i+1}$. 
\end{deff}

\begin{deff}[$\omega$-continuity] $F : \lattice_1 \to \lattice_2$ is $\omega$-continuous if it preserves joins over $\omega$-chains, i.e. for all $(a_n)_{n \geq 0}$:
\[F\left(\bigvee_n a_n\right) = \bigsqcup_n F(a_n) \]
\end{deff} \noindent\\
Notice that continuity implies $\omega$-continuity (the converse is false). 
These next two lemmas will also prove useful in the next chapters.

\begin{lemma}
\label{join_of_monotonic}
 For any two monotonic functions $F$ and $G$, the function: \[H (x) =  (F(x) \vee G(x))\] is also monotonic.
\end{lemma}
%
%
%
%
\medskip
\begin{lemma}
 \label{join_of_omegacont}For any two $\omega$-continuous functions $F$ and $G$, the function: \[H (x)=  (F(x) \vee G(x))\] is also $\omega$-continuous.
\end{lemma}

%
%
%

\section{Adjunctions \& Fixed Point Calculus}
This section summarizes the definitions and main properties of Galois connections and fixed points. From these we define a set of rules (named \emph{Fixed Point Calculus}, see e.g. \cite{mathematics1995fixed}) that we shall employ in the proofs in the following chapters.

\subsection{Galois Connections}

\begin{deff}
 Given two lattices $(\lattice_1, \leq_1)$ and $(\lattice_2, \leq_2)$, we say that a pair of monotonic functions $F:\lattice_1 \to \lattice_2$ and $G:\lattice_2 \to \lattice_1$ forms a \emph{Galois connection} if:
 
 \[ F(x) \leq_2 y \iff x \leq_1 G(y) \]
 
 $F$ is referred to as the \emph{left adjoint} of $G$, while $G$ as the \emph{right adjoint} of $F$. We also write $F \dashv G$.
\end{deff}

\paragraph{Lifting} Given any partial order $\leq \,\,\in Rel (X, X)$, its lifted version $\dot \leq \in Rel (X^X, X^X)$ is defined by:
\[f \,\,\dot\leq\,\, g \iff \text{for all } x \in X \text{ : } f(x) \leq g(x)\]
The proofs of the following two lemmas, that will be used extensively throughout the present manuscript, can be found in \cite{DaveyPriestley/Book/1990}.

\begin{lemma}[Cancellation]\label{cancellation_lemma} Let $(\lattice_1, \leq_1), (\lattice_2, \leq_2)$ be partial orders, and $F:\lattice_1 \to \lattice_2$, $G:\lattice_2 \to \lattice_1$ be monotonic functions. Then $F$ and $G$ form a Galois connection if and only if both the following conditions are satisfied:

\[F \circ G  \dotleq{2} id_2
\quad\quad\quad\quad
id_1 \dotleq{1} G \circ F\]
Where $id_1 : \lattice_1 \to \lattice_1$ and $id_2 : \lattice_2 \to \lattice_2$ are the identity functions on $\lattice_1$ and $\lattice_2$.
\end{lemma}

\begin{lemma}\label{continuous_adjoint}
 Let $(\lattice_1, \leq_1), (\lattice_2, \leq_2)$  be complete lattices, $f : \lattice_1 \to \lattice_2$ be a continuous function, and $g : \lattice_2 \to \lattice_1$ defined as follows, for all $y \in \lattice_2$:
 
 \[ g(y) = \bigvee \{x \mid f(x) \leq_2 y\} \]
 Then $f \dashv g$.
\end{lemma}

\subsection{Fixed Points \& Fixed point calculus}

A fixed point of the function $F: \lattice \to \lattice$ is an element $a \in \lattice$ such that $F(a) = a$. The lattice structure of $\lattice$ makes it possible to define the notions of least fixed point, denoted $\mu x . F(x)$ or $\mu F$, and greatest fixed point, denoted $\nu x . F(x)$ or $\nu F$. The following theorem expresses a sufficient condition for the existence of fixed points of a lattice endofunction, and together with this theorem comes an \emph{induction principle}, which shall be extremely useful in the chapters to come.

 \begin{theo} [Knaster-Tarski]
  Let $F: \lattice \to \lattice$ be a monotonic function on a complete lattice $(\lattice, \leq)$. Then $F$ has a least fixed point $\mu F$, and:
 \[ \mu F = \bigwedge \{x : F(x) \leq x\} 
\]

 \end{theo}


\begin{corollary}As a consequence we get the following \emph{fixed point induction principle}:

 \begin{equation}
 F(x) \leq x \implies \mu F \leq x \label{KT-induction} \tag{KT induction}   
 \end{equation}


\end{corollary}
The definition of $\mu F$ given above is impredicative and non-constructive. A constructive and iterative definition of $\mu F$ can be obtained by assuming the $\omega$-continuity of $F$ (see Theorem \ref{kleene_theorem}). 

\paragraph{Fixed Point Calculus} The next two propositions, along with \eqref{KT-induction} and Lemmas \ref{cancellation_lemma} and \ref{continuous_adjoint}, form the \emph{fixed point calculus} \cite{mathematics1995fixed} we shall employ in proofs throughout the following chapters.

\begin{prop} Let $F, G, H: \lattice \to \lattice$ be monotonic. Then:
 \begin{enumerate}
  \item ($\mu$ \textbf{monotonic}) $F \dotleqx G \implies \mu F \leq \mu G$ 
  \item (\textbf{rolling rule}) $\mu (F \circ G) = F (\mu (G \circ F))$
  \item (\textbf{diagonal rule}) $\mu x . (x \oplus x) = \mu x.\mu y. (x \oplus y)$, for any monotonic binary operator $\oplus$.
  \item (\textbf{simple $\mu$-fusion}) $F \circ G \dotleqx G \circ H \implies  \mu F \leq G(\mu H)$
 \end{enumerate}
\end{prop}

\begin{prop}[\textbf{$\mu$-fusion}] Let $F : \lattice \to \lattice$ be continuous and $G, H: \lattice \to \lattice$ be monotonic. Then:
\begin{enumerate}
 \item $F \circ G \dotleqx H \circ F \implies F(\mu G) \leq \mu H$
 \item $F \circ G = H \circ F \implies F(\mu G) = \mu H$
\end{enumerate}
The proof of this proposition can be found in \cite{mathematics1995fixed}.
 
\end{prop}

Another lemma that will prove very useful in the following chapters is the following.
 \begin{lemma}\label{BonksLemma}
  Let $f, h : \lattice \to \lattice$ be monotonic. Let $G(a) = \mu x . a \vee f(x)$. If $f\circ h \,\,\dot\leq\,\, h\circ f$, then:
\[(G\circ h)(a) \leq (h\circ G)(a)\]

 \begin{proof}
  Let $f\circ h \,\,\dot\leq\,\, f\circ h$. By the definition of fixed point:
  \begin{equation}
  a \vee (f\circ G)(a) \leq G(a)\tag{A}
  \end{equation}
  \begin{align*}
   G(h(a))&= h(a) \vee (f \circ h \circ G)(a) \\
   &\leq h(a) \vee (h \circ f \circ G)(a) \tag{Hp and monotonicity of $\vee$}\\ &\leq
   h(a \vee (f \circ G)(a)) \tag{monotonicity of $h$ and Remark \ref{lax_pres_join}}\\
   &\leq (h \circ G)(a) \tag{lax continuity of $h$ and A}
  \end{align*}
 \end{proof}
 
 \subsection{Fixed Points \& Iteration}
\begin{theo}[Kleene] \label{kleene_theorem}
 
 Let $F$ be a monotonic and $\omega$-continuous function on a complete lattice $(\lattice, \leq)$. Then: \[\mu F = \bigvee_{n \geq 0} F^n(\perp)\]

\end{theo}
Where $F^n$ is defined inductively as:
 \begin{itemize}
 \item $F^0(x) = x$
 \item $F^{n+1} (x) = F(F^n(x))$
 \end{itemize}

\begin{corollary}[$\omega$-continuous fixed point induction, \cite{Gavazzo/LICS/2023}]\label{kleene-induction}
Let $F$ be $\omega$-continuous:
 \[ (x \leq \mu F \wedge a \implies F(x) \leq a) \implies \mu F \leq a \]
For any $x$.
\end{corollary}

 \end{lemma}

\begin{lemma}[Enhanced $\omega$-continuous fixed point induction, \cite{Gavazzo/LICS/2023}] \label{ehanced-kleene-ind}
 Let $F$, $G : \lattice \to \lattice$ be $\omega$-continuous, and $G$ be strict. Then, to prove $G(\mu F) \leq a$ it is sufficient to show that for any $x$:
 \[ x \leq \mu F \,\,\&\,\, G(x) \leq a \implies G(F(x)) \leq a \]
\end{lemma}

\section{Algebra of Relations}
\begin{figure}
  \centering
  \footnotesize
\begin{framed}
     \begin{subfigure}[b]{0.3\textwidth}
         \centering
         \begin{align*}
            \bot &\leq a
            \\
            a &\leq \top
            \\
            \bigvee a_i \leq b &\iff
            \forall i.\ a_i \leq b
            \\
            b \leq \bigwedge a_i &\iff \forall i.\ b \leq a_i
            \\
             a; \Delta &= a
            \\
            \Delta;a &= a
            \\
            a; (b; c) &= (a; b);c
        \end{align*}
         \caption{Lattice and monoid laws}
         \label{fig:algebraic-laws-lattice}
     \end{subfigure}
     \hfill
     \begin{subfigure}[b]{0.4\textwidth}
         \centering
            \begin{align*}
                a;\bot &= \bot 
                \\
                \bot; a &= \bot
                \\
                a; \Big(\bigvee_i b_i\Big) &= \bigvee_i (a;b_i)
                \\
                \Big(\bigvee_i a_i \Big); b &= \bigvee_i (a_i;b)
                \\
                a_1 \leq b_1 \,\,\&\,\, a_2 \leq b_2 
                &\implies a_1;a_2 \leq b_1;b_2
            \end{align*}         
        \caption{Quantale Laws}
         \label{fig:algenbraic-laws-quantale}
     \end{subfigure}
     \hfill
     \begin{subfigure}[b]{0.2\textwidth}
         \centering
            \begin{align*}
            a\op \leq b \iff& a \leq b\op
            \\
            \Delta\op &= \Delta
            \\
            \bot\op &= \bot
            \\
            \top\op &= \top
            \\
            a\op &= a
            \\
            (a;b)\op &= b\op; a\op
            \\
            \Big(\bigvee a_i\Big)\op &= \bigvee a_i\op
        \end{align*}
        \caption{Converse Laws}
         \label{fig:algebraic-laws-converse}
     \end{subfigure}
\end{framed}
        \caption{Algebraic Laws}
        \label{fig:algebraic-laws}
\end{figure}

This section defines and summarizes the main properties of \emph{relation algebras}. For a full account on the topic, see \cite{relational-mathematics}. Relation algebras allow us to express several mathematical concepts (e.g. partial and total functions); in this manuscript we shall employ it mainly to reason about certain properties of relations, which shall be mentioned in \ref{subs:alg_orders}.

\begin{deff}
\begin{itemize}
 \item []
 \item A (unital) \emph{quantale} \cite{Rosenthal/Quantales/1990} $(A, \leq)$ is a complete lattice with an additional join preserving operation $\tt ;$ (\emph{sequential composition}) such that $(A, \tt ;)$ forms a monoid with unit $\Delta$.
 \item Given a quantale, an involution on it is an operator $-\op$ such that:
  \[ a^\circ \leq b \iff a \leq b^\circ \quad\quad \Delta\op = \Delta\quad\quad (a; b)\op = b\op ; a\op\]
  
 \end{itemize}

 \paragraph{Modular law} This law, also known as ``Dedekind Law'', acts as an interface between all three structures:
\[ (a;b) \wedge c \leq (a \wedge c;b^\circ); b \]
This rule plays a key role in proving many useful properties of relations. However, understanding
such a rule is not straightforward, and thus we will make any use of it (which will be minimal for
our purposes) explicit. 

\begin{deff}
 An algebra of relations $(\ralg, \leq, ;, \Delta, -^\circ, \vee, \wedge)$ is an involutive, unital quantale satisfying the modular law.
\end{deff}
 
\end{deff}

  \begin{example}
   
In terms of binary relations in $\Rel A$, the complete lattice structure is given by the subset inclusion $\subseteq$. The join and meet are therefore given by the set-theoretic union and intersection, the top and bottom relations are respectively $A \times A$ and $\emptyset$. The identity $\Delta$ is the usual identity:
\begin{equation}
    \rel x \Delta y \iff \rel x = y \label{identity_deff} \tag{def. $\Delta$}
\end{equation}
The relation $;$ is the usual composition:
\begin{equation}
 \rel x {(a ; b)} y \iff \exists z. (\rel x a z \text{ and } \rel z b y) \label{composition_deff}
\tag{def. $;$}
\end{equation}
 which is associative and has a unit $\Delta$ such that $x \Delta y$ if and only if $x = y$. Additionally, sequential composition is join preserving:
 \begin{equation}
 a; \Big(\bigvee_i b_i\Big) = \bigvee_i (a;b_i)
                \quad\quad
                \Big(\bigvee_i a_i \Big); b = \bigvee_i (a_i;b)
  \label{composition_cont}
\tag{cont. $;$}
\end{equation}
 and therefore monotonic w.r.t. $\leq$:
\begin{equation}
a_1 \leq b_1 \,\,\&\,\, a_2 \leq b_2 \implies a_1;a_2 \leq b_1;b_2
  \label{composition_monot}
\tag{monot. $;$}
\end{equation}

 The involution operation $-^\circ$, referred to as \emph{converse}, is an operator obeying the following rule:
 \begin{equation}
     a^\circ \leq b \iff a \leq b^\circ \tag {monot. $-\op$ } \label{op_rule_deff}
 \end{equation}
  The converse operator on $\Rel A$ is given by the inverse of a relation, i.e. for all $a \in \Rel A$ and ${\tt x, y} \in A$:
  \begin{equation}
      \rel y {a\op} x \iff \rel x a y \label{op_deff} \tag{def. $-\op$}
  \end{equation}
This operator is its own inverse: 
\begin{equation}
    {a\op}\op = a \label{op_involution} \tag{invol. $-\op$}
\end{equation}
it is continuous, and is related with composition by the following:
\begin{equation}
(a ; b)\op = b\op ; a\op \label{op_composition_rule}
\end{equation} 
  \end{example}

The algebraic laws that relate the operators of an algebra of relations are found in Figure \ref{fig:algebraic-laws}.

\begin{deff}
 A \emph{morphism} of algebras of relations $\ralg$, $\ralgb$ is a function $F: A \to B$ such that:
\[F(\Delta) = \Delta \quad\quad F(a ; b) = F(a) ; F(b) \quad\quad F(a^\circ) = F(a)^\circ\]
We say that $F$ is monotonic or ($\omega$-)continuous if it is so on the underlying lattice structure.
\end{deff}

Notice that since algebras of relations are, in particular, complete lattices, 
we can rely on Lemma~\ref{continuous_adjoint} to prove the existence of adjoints of 
continuous functions. For instance, since composition $;$ preserves joins in both 
arguments, Lemma~\ref{continuous_adjoint} entails the 
existence of two binary operators $\backslash$ and $/$ 
defined by the rules:
\begin{align}
    a;b \leq c & \iff a \leq c / b
    \label{comp-adj-right}
    \tag{$/$-def}
    \\
    a;b \leq c & \iff b \leq a \backslash c
    \label{comp-adj-left}
    \tag{$\backslash$-def}
\end{align}
As an immediate consequence of these laws, we obtain their associated cancellation laws, which 
will be useful in some calculations. 
\begin{align}
    (a / b); b &\leq a
    \label{comp-adj-cancellation-right}
    \tag{$/$-cancellation}
    \\
    a;(a\backslash b) &\leq b 
    \label{comp-adj-cancellation-left}
    \tag{$\backslash$-cancellation}
\end{align}



 

\subsection{Algebra of orders}\label{subs:alg_orders}
Algebras of relations allow us to reason about properties of relations such as reflexivity, symmetry and transitivity. Fixed an algebra of relations $\ralg$:

\begin{prop} A relation $a$ is:
\begin{itemize}
 \item reflexive if and only if $\Delta \leq a$ 
 \item transitive if and only if $a;a \leq a$
 \item symmetric if and only if $a\op \leq a$
\end{itemize}
\end{prop}

We can define closure operators for these properties.

\begin{deff}
 A closure operator is a monotone, idempotent function $C: \ralg \to \ralg$ such that, for all $a$, $a \leq C(a)$.
\end{deff}

In general, the closure of a relation $a$ with respect to some property is the least (w.r.t. $\leq$) relation that satisfies such property. In particular, the closure operators for the properties mentioned above are defined as follows:
\begin{deff} 
 \begin{itemize}
 \item []
  \item The reflexive closure of a relation $a$ is $a^{\mathit{refl}} := a \vee \Delta$.
  \item The symmetric closure of a relation $a$ is $a^{\mathit{sym}} := a \vee a\op$.
  \item The transitive closure is slightly more complex to define. It is defined as follows: $a^+ := \mu x . a;x$.
 \end{itemize}

 Additionally, it is extremely useful to define the \emph{transitive and reflexive} closure, often called the \emph{Kleene star}. This operator models reachability through a finite number of steps (possibly zero), and will have a central role in the following chapters.
 
\end{deff}

\begin{deff}
 The Kleene star of a relation $a$ is $a^* = \mu x . \Delta \vee a;x$. 
 
Let $F(x) = \Delta \vee a;x$. This function can be shown to be $\omega$-continuous. Hence by Kleene's theorem:
 \[ a^* =  \bigvee_{n \geq 0} F^n (\perp)\]
\end{deff}
\begin{remark}
It can be seen rather easily that $F^n$ is simply $a^{(n)} := \underbrace{a ; \hdots ;a}_n$. Because of this, the following holds:
\begin{equation}
    a^{(n)} \leq a^* \label{a^n-in-astar} \tag{$x^{(n)} \leq x^*$}
\end{equation}
\end{remark}\noindent
We state the closure properties of the Kleene star explicitly in the following proposition.

\begin{prop} The Kleene star is a closure operator, that is, it is monotonic:
\begin{equation}
    a \leq b \implies a^* \leq b^* \label{kleene-monot} \tag{Kleene monot.}
\end{equation}
idempotent:
\begin{equation}
    a^{**} = a^* \label{kleene-idemp} \tag{Kleene idemp.}
\end{equation}
and such that:
\begin{equation}
    a \leq a^* \label{kleene-a-in-astar} \tag{$x \leq x^*$}
\end{equation}
\end{prop}\noindent
The Kleene star has an additional property that relates it to $-\op$:
\begin{prop} For all relations $a \in \ralg$, the following holds: 
\begin{equation}
{a^*}\op = {a\op}^* \label{kleene-op}\tag{${x^*}\op = {x\op}^*$}
\end{equation}
\end{prop}

\paragraph{Coreflexives}
Algebras of relations also allow us to reason about (unary) properties, so that we can 
recover the so-called calculus of classes inside them. The key notion is the one of a 
\emph{coreflexive}~\cite{algebra-of-programming,scedrov-freyd}. 

\begin{deff}
    A relation $a$ is a \emph{coreflexive} if $a \leq \Delta$.
\end{deff}

Thinking in terms of binary relations on a set $A$, we see that a coreflexive is a subset of the identity 
relation; consequently a coreflexive $a$ identifies a subset $\alpha$ of $A$, viz. $\alpha := \{{\tt a} \mid 
{\tt a} \mathrel{a} {\tt a}\}$. Vice versa, any $\alpha \subseteq A$ identifies a coreflexive 
$a :=  \{({\tt a}, {\tt a}) \mid {\tt a} \in \alpha\}$. 

We will use coreflexives to encode subsets and properties in algebras of relations. For instance, in our relational 
theory of rewriting, we shall
extensively use coreflexives identifying variables and constants. Notice that coreflexives are self-dual, meaning 
that $a^{\circ} = a$, and the monoid structure collapses to the lattice one when restricted to coreflexives. 
In particular, for coreflexives $a,b$, we have $a;b = a \wedge b$, so that relation composition becomes 
commutative and idempotent on coreflexives.

\chapter{Abstract Rewriting, Relationally}
\label{chapter:abstract_rewriting}
\section{Abstract Reduction Systems}
Rewriting theory studies discrete transformations between expressions. These expressions may be, for example, computer programs, arithmetical expressions, $\lambda$-terms or graphs. The common substratum of rewriting theory, that is, the underlying structure that these concrete examples have in common, is studied through \emph{Abstract Reduction Systems} (ARSs). We first present ARSs traditionally, and then show that it is possible to study abstract rewriting properties through algebras of relations. Properties like the ones discussed in this section have been studied through formalisms similar to the one we are using here, e.g. using Kleene Algebras \cite{struth-1, Struth-abstract-abstract-reduction}.

\begin{deff} [ARS, traditionally]
  An ARS is given by a set $A$ together with a relation $\to \,\, \subseteq A \times A$, called \emph{reduction relation}.
 \end{deff}
 
Since we are interested in studying abstract reductions systems using algebras of relations, we need to define a \emph{relational notion} of ARS \cite{baumer}.
 
 \begin{deff} [ARS, relationally] Fixed an algebra of relations $\ralg$, an $\ralg$-ARS (or \emph{relational} ARS) is simply a relation $a \in \ralg$.
 \end{deff}

  \begin{example} Given a set $A$ and $\Rel A$ as a relation algebra, then an ARS is a binary endorelation on $A$. 
  \end{example}
  Notice that any ARS $(A, \to)$ gives an $\ralg$-ARS by taking $\ralg = \Rel A$.
  
  \begin{notation}
   Traditionally, the inverse of a relation $\to$ is denoted $\leftarrow$, its reflexive closure is denoted $\to^\equiv$, its symmetric closure $\leftrightarrow$ and its transitive, reflexive closure $\to^*$. Figure \ref{tab1::fig} shows the corresponding notation for $\ralg$-ARSs.
  \end{notation}

  \begin{figure}[t]
  \begin{center}
   \begin{tabular}{|c|c|c|c|c|}
   \hline
    $\to$ & $\leftarrow$ & $\leftrightarrow$ &$\to^\equiv$ & $\to^*$\\
    \hline
    $a$ & $a\op$ & $a \vee a\op$ &$a \vee \Delta$ & $a^*$\\
    \hline
   \end{tabular}
  \end{center}
   \caption{Abstract rewriting notation and its relational counterpart}
   \label{tab1::fig}
  \end{figure}

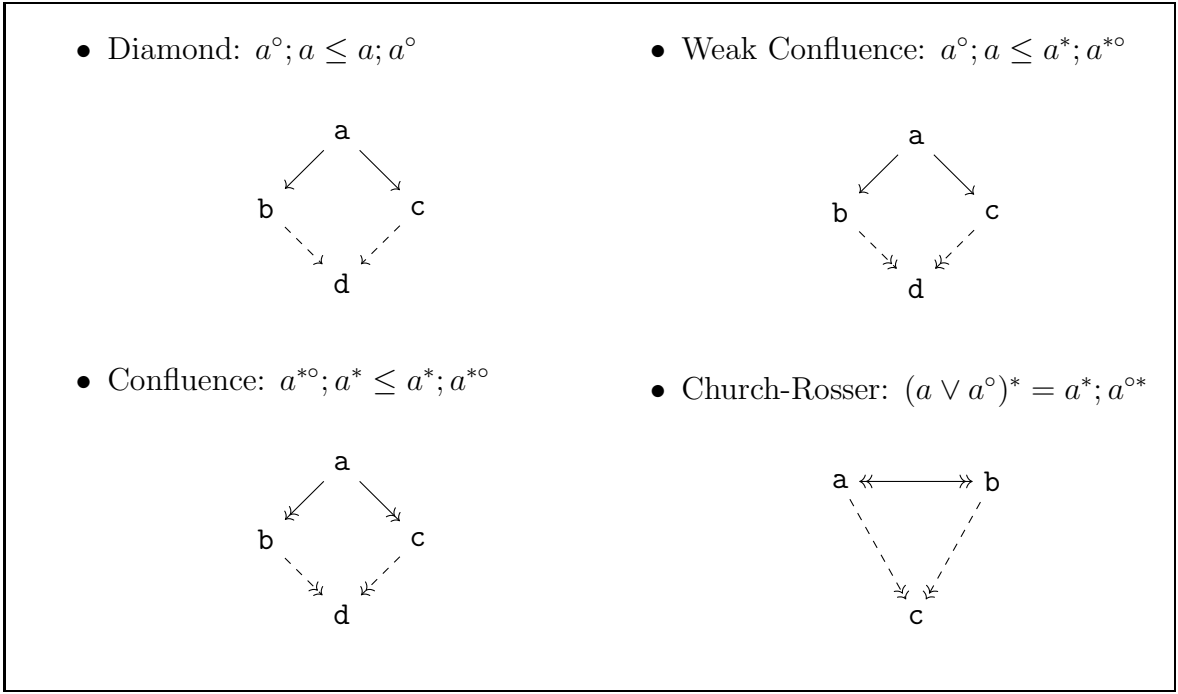
\begin{figure}
\begin{framed}
 \begin{multicols}{2}
  \begin{itemize}
   \item Diamond: $a\op ; a \leq a ; a\op$
   
   \begin{center}
   \begin{tikzpicture}
    \node (A) at (0,1) {\tt a};
    \node (B) at (-1,0) {\tt b};
    \node (C) at (1,0) {\tt c};
    \node (D) at (0,-1) {\tt d};
    
    \draw[->] (A) -- (B);
    \draw[->] (A) -- (C);
    \draw[->, dashed] (B) -- (D);
    \draw[->, dashed] (C) -- (D);
   \end{tikzpicture}
   \end{center}

   \item Confluence: $a^{*\circ} ; a^{*} \leq a^{*} ; a^{*\circ}$
   
   \begin{center}
   \begin{tikzpicture}
    \node (A) at (0,1) {\tt a};
    \node (B) at (-1,0) {\tt b};
    \node (C) at (1,0) {\tt c};
    \node (D) at (0,-1) {\tt d};
    
    \draw[->>] (A) -- (B);
    \draw[->>] (A) -- (C);
    \draw[->>, dashed] (B) -- (D);
    \draw[->>, dashed] (C) -- (D);
   \end{tikzpicture}
   \end{center}
   
   \item Weak Confluence: $a\op ; a \leq a^{*} ; a^{*\circ}$
   
    \begin{center}
   \begin{tikzpicture}
    \node (A) at (0,1) {\tt a};
    \node (B) at (-1,0) {\tt b};
    \node (C) at (1,0) {\tt c};
    \node (D) at (0,-1) {\tt d};
    
    \draw[->] (A) -- (B);
    \draw[->] (A) -- (C);
    \draw[->>, dashed] (B) -- (D);
    \draw[->>, dashed] (C) -- (D);
   \end{tikzpicture}
   \end{center}
   
   \item Church-Rosser: $(a\vee a\op)^{*} = a^{*} ; a^{\circ*}$
   
    \begin{center}
   \begin{tikzpicture}
    \node (A) at (-1,1.8) {\tt a};
    \node (B) at (1,1.8) {\tt b};
    \node (C) at (0,0) {\tt c};
    
    \draw[<<->>] (A) -- (B);
    \draw[->>, dashed] (A) -- (C);
    \draw[->>, dashed] (B) -- (C);
   \end{tikzpicture}
   \end{center}
  \end{itemize}

  \end{multicols}  
\end{framed}
\caption{Some properties of rewriting systems, expressed relationally.}
\label{rewriting_properties::fig}
\end{figure}

  \paragraph{Relations at work: confluence} Fixed an ARS $(A, \to)$, one may define and study algebraically several interesting rewriting properties, such as the ones summarized in Figure \ref{rewriting_properties::fig}. To showcase this approach, and give an example of how it is possible to define and prove theorems on well-known rewriting properties using the tools of relation algebra, we shall now define the \emph{confluence} property, first in the usual way and then relationally. We will then show the equivalence of these definitions, and prove a well-known theorem using it.

  \begin{deff}[Confluence] Given an ARS $(A, \to)$:
  \begin{itemize}
   \item ${\tt t} \in A$ is said to be \emph{confluent} if for all $\tt s$, ${\tt r} \in A$ such that $\rel t {\to^{*}} s$ and $\rel t {\to^{*}} r$, there exists ${\tt x} \in A$ such that $\rel s {\to^{*}} x$ and $\rel r {\to^{*}} x$.
   \item A reduction $\to$ is confluent if all ${\tt t} \in A$ are confluent.
  \end{itemize}
  \end{deff}

  \begin{prop} [Confluence, relationally]
   Given an ARS $(A, \to)$ and the $\ralg$-ARS $a$ such that $a = \,\,\to$, the reduction $a$ is confluent if and only if $a^{*\circ} ; a^{*} \leq a^{*} ; a^{*\circ}$.
  \end{prop}
  \begin{proof}
   The statement $a^{*\circ} ; a^{*} \leq a^{*} ; a^{*\circ}$ is logically equivalent to:
   \[\forall {\tt s}, {\tt r} \in A . \rel s{(a^{*\circ} ; a^{*})}r \implies \rel s{(a^{*} ; a^{*\circ})}r \]
  Note that the left part of this implication is, by definitons of composition and opposite, equivalent to: \[\exists {\tt t} \in A .\,\, \rel t {a^{*}} s \logicand \rel t {a^{*}} r \]
  while the right part is equivalent to:
  \[\exists {\tt x} \in A .\,\, \rel s {a^{*}} x \logicand \rel r {a^{*}} x \]
  Putting these two together, we obtain that $a^{*\circ} ; a^{*} \leq a^{*} ; a^{*\circ}$ if and only if:
  \[\forall {\tt s}, {\tt r} \in A.\,\, \exists {\tt t} \in A .\,\, \rel t {a^{*}} s \logicand \rel t {a^{*}} r \implies \exists {\tt x} \in A .\,\, \rel s {a^{*}} x \logicand \rel r {a^{*}} x \]
  By FOL, if $x \notin FV(\beta)$ we have:
  \[(\exists x .\,\, \alpha) \implies \beta \equiv \forall x .\,\, (\alpha \implies \beta)\]
  Hence
  \[\forall {\tt t}, {\tt s}, {\tt r} \in A.\,\, \rel t {a^{*}} s \logicand \rel t {a^{*}} r \implies \exists {\tt x} \in A .\,\, \rel s {a^{*}} x \logicand \rel r {a^{*}} x \]
  Which is the definition of confluence given above.
  \end{proof}
  
  We now give the definition of another important property of ARSs, the Church-Rosser property. It is well known that such property is in fact equivalent to confluence, and we provide a proof of this theorem with the goal of showcasing the relational approach.
  
  \begin{deff} Given an ARS $(A, \to)$, we say that 
  $\to$ has the \emph{Church-Rosser} property if 
  for all $\tt x$, ${\tt y} \in A$ such that $\rel x {\overset{*}\leftrightarrow} y$ there exists ${\tt z} \in A$ such that $\rel x {\to^{*}} z$ and $\rel y {\to^{*}} z$. 
  \end{deff}
  It is easy to realize that the natural relational counterpart of the Church-Rosser property 
  is given by the following inequality: 
  \begin{equation}
      (a\vee a\op)^{*} = a^{*} ; a^{\circ*}
      \label{eq-CR}
      \tag{CR}
  \end{equation}
Indeed, when instantiated on concrete ARSs, \eqref{eq-CR} precisely gives the usual 
Church-Rosser property. 

  \begin{prop}[Church-Rosser, Relationally]
      Given an ARS $(A, \to)$ and the $\ralg$-ARS $a$ such that $a = \,\,\to$, the reduction $a$ has 
      the Church-Rosser property if and only 
     \eqref{eq-CR} holds.
  \end{prop}\noindent
If the reader understood the previous proof, the validity of the above proposition shall be evident. 
  
In light of the previous results, when working with $\ralg$-ARSs we freely 
use expressions such as \emph{confluent}, \emph{Church-Rosser}, $\hdots$ 
as synonyms of their relational (and algebraic) counterparts. 

\pagebreak

  \begin{theo} 
   A relation $\relone \in \ralg$ is confluent precisely when it is Church-Rosser.
   \begin{proof}
    \begin{enumerate}
    \item []
    \item
    We first prove that the Church-Rosser property \eqref{eq-CR} 
    implies confluence. Let $a \in \ralg$ 
    and assume \eqref{eq-CR}.
    To prove confluence, we first notice that 
    \begin{align*}
        a^{*\circ};a^* \leq a^*;a^{*\circ} 
        &\iff a^{*\circ} \leq (a^*;a^{*\circ})/a^*
        \tag{\ref{comp-adj-right}}
        \\
        &\iff a^{\circ*} \leq (a^*;a^{*\circ})/a^*
        \tag{\ref{kleene-op}}
        \\
        &\iff  \mu x. \idrel \vee a^{\circ};x \leq (a^*;a^{*\circ})/a^* 
        \tag{\text{Definition of Kleene star}}
        \\
        &\impliedby \Delta \vee (a^\circ;((a^*;a^{*\circ})/a^*)) \leq 
        (a^*;a^{*\circ})/a^* 
        \tag{\ref{KT-induction}}
    \end{align*}
    It is thus sufficient to prove 
    \begin{align}
        \Delta &\leq (a^*;a^{*\circ})/a^*
        \label{CR-implies-confluence-1}
        \\
        a^\circ;((a^*;a^{*\circ})/a^*) &\leq 
        (a^*;a^{*\circ})/a^*
        \label{CR-implies-confluence-2}
    \end{align}
For \eqref{CR-implies-confluence-1}, we have:
\begin{align*}
    \Delta \leq (a^*;a^{*\circ})/a^* 
    &\iff 
    a^* \leq a^*;a^{*\circ}
    \tag{\ref{comp-adj-right}}
    \\
    &\iff 
    a^*; \Delta \leq a^*;a^{*\circ}
    \tag{$x = x;\Delta$}
    \\
    &\impliedby \Delta \leq a^{*\circ}
\end{align*}
For \eqref{CR-implies-confluence-2}, we have:
\begin{align*}
    a^\circ;((a^*;a^{*\circ})/a^*) \leq (a^*;a^{*\circ})/a^*
    &\iff 
    a^\circ;((a^*;a^{*\circ})/a^*);a^* \leq a^*;a^{*\circ}
    \tag{\ref{comp-adj-right}}
    \\
    &\impliedby 
    a^\circ;a^*;a^{*\circ} \leq a^*;a^{*\circ}
    \tag{\ref{comp-adj-cancellation-right} \text{ on } $(a^*;a^{*\circ})/a^*);a^*$}
    \\
    &\iff
    a^\circ;(a \vee a\op)^{*} \leq a^*;a^{*\circ}
    \tag{\ref{eq-CR}}
    \\
    &\impliedby
    (a \vee a^\circ);(a \vee a\op)^{*} \leq a^*;a^{*\circ}
    \tag{$a\op \leq a \vee a\op$}
    \\
    &\impliedby
    (a \vee a\op)^{*} \leq a^*;a^{*\circ}
    \tag{$x;x^* \leq x^*$}
    \\
    &\impliedby
    \eqref{eq-CR}
\end{align*}

     \item We now show that confluence implies CR. Assume $a^{*\circ} ; a^{*} \leq a^{*} ; a^{*\circ}$. In order to prove that $a$ is CR we need to show the following inequalities:
     \begin{equation}
     a^*; {a\op}^* \leq (a \vee a\op)^* \label{CRconfeq1}
     \end{equation}
     \begin{equation}
(a \vee a\op)^* \leq a^*; {a\op}^*\label{CRconfeq2}
     \end{equation}
     
     \begin{enumerate}
      \item [\eqref{CRconfeq1}] By the monotonicity of the Kleene star the following holds: \[a^*; {a\op}^* \leq (a \vee a\op)^* ; (a \vee a\op)^*\]
      And by $R^*;R^* \leq R^*$ we conclude: $a^*; {a\op}^* \leq (a \vee a\op)^*$
      \item [\eqref{CRconfeq2}] By fixed point induction \eqref{KT-induction}, as $b^* = \mu x .\Delta \vee b;x$, we have:
      \[ \Delta \vee (a \vee a\op); a^*; {a\op}^* \leq a^*; {a\op}^* \implies (a \vee a\op)^* \leq a^*; {a\op}^* \]
      Firstly we have $\Delta \leq \Delta ; \Delta \leq a^* ; {a\op}^*$ hence we only need to prove:
      \[ (a \vee a\op); a^*; {a\op}^* \leq a^*; {a\op}^* \]
      Which can be proven as follows:
      \begin{align*}
       (a \vee a\op); a^*; {a\op}^* &= a ; a^*; {a\op}^*  \vee a\op ; a^*; {a\op}^* \\
       &\leq a^*; {a\op}^*  \vee a\op ; a^*; {a\op}^* \tag{$a; a^* \leq a^*$ and \eqref{composition_monot}} \\
       &\leq a^*; {a\op}^*  \vee {a\op}^* ; a^*; {a\op}^* \tag{$a\op \leq {a\op}^*$ and \eqref{composition_monot}} \\
       &\leq a^*; {a\op}^*  \vee {a\op}^* ; a^*; {a\op}^* \tag{Hp. confluence} \\
       &\leq a^*; {a\op}^*  \vee {a}^* ; {a\op}^*; {a\op}^* \tag{Hp. confluence} \\
       &\leq a^*; {a\op}^*  \vee {a}^* ; {a\op}^* \tag{$b \vee b = b$ } \\
       &=  a^*; {a\op}^*
      \end{align*}

     \end{enumerate}
     \end{enumerate}
   \end{proof}

  \end{theo}

\subsection{Beyond Confluence} \label{subsec:beyond_conf}
At this point, the reader should be reasonably convinced of 
the effectiveness of the relational approach in the study of 
abstract rewriting systems, at least for what concerns 
confluence-like properties. Even if we do not investigate them 
in this manuscript, it is also worth mentioning that several other 
rewriting properties have been successfully studied in a relational setting, 
notable examples of those being the study of 
termination  -- where, in particular, Doornobos et al. have given 
a relational proof of Newman's Lemma~\cite{backshouse-calculational-approach-to-mathematical-induction}, 
-- and the theory of abstract reduction in the setting of (modal) Kleene algebras by 
Struth~\cite{Struth-abstract-abstract-reduction}.

\chapter{Term Based Rewriting, Relationally}
\label{chapter:term_rewriting}
A \emph{term rewriting system} (TRS) is an ARS such that the objects are \emph{first order terms} and the reduction relation is presented as a set of \emph{ground rewrite rules}. Unlike for ARSs, these rewrite rules do not express the whole reduction relation; they only serve as base cases, from which we build the whole relation by closing under certain properties. For example, we may define a simple system of arithmetic as follows. 

\begin{example}
 \label{arithmetic_example}
Let the terms be defined by the following grammar\footnote{E.g. the natural number $3$ is represented as $\tt S(S(S(0)))$, the expression $2+1$ is represented as $\tt A(S(S(0)), S(0))$, and the expression $0 \times (1 + 0)$ is represented as: $\tt M(0, A(S(0)), 0)$.}:
\[\tt t, s ::= x \mid 0 \mid S(t) \mid A(t, s) \mid M(t, s)\]
In this case the \emph{ground} rewrite rules we may define could be the following:

\begin{multicols}{2}

\begin{center}

$ \tt A(0, x) \mapsto x $

$ \tt A(S(x), y) \mapsto S(A(x, y))$

$ \tt M(0, x) \mapsto 0 $

$ \tt M(S(x), y) \mapsto A(M(x, y), y) $

\end{center}

\end{multicols}

\end{example}
In this chapter we begin to showcase a relational formalization of TRSs, originally introduced in \cite{Gavazzo/LICS/2023}, which we shall build on in the next chapter. Algebras of relations are enough to express the properties of ARSs, but when it comes to syntax, they are not sufficient anymore \cite{ghani-luth-coinserters}. It is therefore necessary to enrich algebras of relations with specially defined operators that allow us to reason about syntax.

\section{TRSs: Basic Definitions}
Before defining the terms themselves, we need a way to formally specify which building blocks they are made of. In particular, a term consists of either a variable or an operator applied to a specified number of other terms. Such operators are defined through signatures.
\begin{deff}[Signature]
 A \emph{signature} $\Sigma$ consists of a set of operator symbols, each with a fixed arity. The arity of an operator symbol is a natural number, representing the number of arguments that it is supposed to have. The \emph{nullary} operators, that is, the ones with arity zero, are usually referred to as \emph{constants}. We denote the set of $n$-ary operator symbols as $\Sigma_n$.
\end{deff}
Now we can formally define terms and rewrite rules.

\begin{deff}[$\Sigma$-terms] 
Given a countable set of variables $V$ and a signature $\Sigma$, we define the set $\termset$ of $\Sigma$-terms over $V$ recursively by the following grammar:
\[ \tt t ::= x \mid {\it o}(t, \hdots, t) \text{\quad\quad where ${\tt x} \in V$ and $o \in \Sigma$.}\]

\end{deff}

\paragraph{Contexts}
When reasoning about terms, it is often convenient to rely on \emph{contexts}, i.e. terms with a linear \emph{hole}. More formally, a hole may be defined as an occourrence of a special symbol $\square$.

\begin{deff}[Context]
 A (linear) \emph{context} is a term over the alphabet $\Sigma \cup \{\square : 0\}$, containing precisely one instance of a special symbol $\square$, which denotes a hole:
 \[ {\tt c} ::= \square \mid o({\tt t_1 \hdots c \hdots t}) \]
\end{deff}

\begin{notation} If $\tt c$ is a context, we represent the term obtained by substituting $\square$ with a subterm $\tt t$ as $\tt c[t]$.
 
\end{notation}

\begin{example}
 Considering Example \ref{arithmetic_example}, the signature is:
 \[ \Sigma = \{ {\tt 0} : 0, {\tt S} : 1, {\tt A} : 2, {\tt M} : 2 \} \]
 In particular, the only constant is $\tt 0$.
 The following is a context:
 \[ \tt c = A(\square, M(S(0), 0)), \]
 and the term $\tt t = A(0, M(S(0), 0))$ is $\tt c[0]$, the term obtained by substituting $\square$ with $\tt 0$.
\end{example}

\paragraph{Substitution}
Substitution is the operation of substituting variables with terms. In the following assume a signature $\Sigma$ to be given.

\begin{deff}
 A \emph{substitution} is a map $\sigma : V \to \termset$.
 
 Given a substitution $\sigma$, we can define a map $\subst \sigma - : \termset \to \termset$ such that:
 \[\subst \sigma x \definedas \sigma(x) \quad\quad  \subst \sigma {o({{\tt t_1}, \hdots, {\tt t_n}})} \definedas o(\subst \sigma {\tt t_1}, \hdots, \subst \sigma {\tt t_n}) \]
 for every $n$-ary operator $o$, i.e. it is a \emph{syntax preserving} map. 
\end{deff}

\begin{notation}
 Given a term $\tt t$ with variables $\tt x_1 \hdots x_n$ and a substitution $\sigma$ such that $\sigma ({\tt x_i}) = {\tt v_i}$, we sometimes write the term $\subst \sigma {\tt t}$ as $\tt t[v_1 \hdots v_n / x_1 \hdots x_n]$.
\end{notation}
%
\subsection{TRSs}
We have defined terms, but we are still to define rewrite rules:
\begin{deff}\label{def:rewritingrule} A \emph{rewrite rule} (or reduction rule) is a pair ${\tt (t, s)} \in \termset \times \termset$ such that $\tt t$ is not a variable.
\end{deff}

\begin{deff} [TRS] A \emph{term rewriting system} (shortly TRS) is a triple $(\Sigma, V, \mapsto)$ where $\Sigma$ is a signature, $V$ is a set of variables and $\mapsto$ is a set of rewriting rules over $\termset$.
\end{deff}

\begin{remark}
The set of rewriting rules $\mapsto$ can be regarded as a relation in $\Rel \termset$, hereafter referred as \emph{ground rule}.
However, not all relations in $Rel (\termset, \termset)$ are sets of rewriting rules since, by Definition \ref{def:rewritingrule}, the pairs  ${\tt (x, s)} \in \termset \times \termset$ where ${\tt x}$ is a variable are not allowed.
\end{remark}

%
%
%

Rewrite rules do not model actual reductions in a TRS, as they model \emph{ground reductions} only. To obtain a useful notion of reduction, one has to extend ground reduction by specifying (a) how the former can be instantiated on arbitrary terms and (b) how reduction is propagated along term constructs. 

For (a), it is enough to close
ground reduction under \emph{substitution instances}: whenever $\tt t$ reduces to $\tt s$ according to a reduction rule, then $\tt \subst \sigma t$
reduces to $\tt \subst \sigma s$. In light of the relational theory of TRSs we shall introduce, it is convenient to slightly depart from traditional exposition of TRSs and to introduce the relation $\trianglearrow$, called \emph{ground reduction instances}, that closes ground reductions under substitution instances.

\begin{deff}
Given a TRS $(\Sigma, V, \mapsto)$, the relation $\trianglearrow$ on $\Sigma$-terms is defined as
 \[\dfrac{\rel t \mapsto s}{ \rel {\subst \sigma t} \trianglearrow {\subst \sigma s}}\]
where $\sigma\colon V \to \termset$ ranges over substitutions.
\end{deff}

For (b), there is more freedom in the design. Indeed, we can postulate that reduction is linear (in the sense
of logic) and sequential, so that all reductions take the form $\tt C[\subst \sigma t] \to C[\subst \sigma s]$, for a reduction rule $\tt t \mapsto s$ (meaning that we
apply exactly one instance of a reduction rule instance in a subterm).

\begin{deff}
\label{def_seqred}
Given a TRS $(\Sigma, V, \mapsto)$, \emph{sequential reduction} is the relation $\to \,\, \subseteq \termset \times \termset$ defined as:
\[ \dfrac{{\rel t \trianglearrow s}}{C[{\tt t}] \to C[{\tt s}]}\]
where $C$ ranges over linear contexts. Equivalently:
\[ \dfrac{{\rel t \mapsto s}}{C[{\tt \subst \sigma t}] \to C[{\tt \subst \sigma s}]}\]
\end{deff}

Definition \ref{def_seqred} is somehow standard in literature of TRSs. For our purposes, however, it is more convenient to work right from the beginning with an inductive characterization of sequential reduction.

\begin{prop}
 For a TRS $(\Sigma, V, \mapsto)$, the relation $\to$ has the following inductive characterization:
 \[ \dfrac{\rel t \trianglearrow s}{\rel t\to s} \quad\quad \dfrac{\rel t \to s \quad o \in \Sigma_n, n \geq 1}{o({\tt \hdots t \hdots}) \to o({\tt \hdots s \hdots})} \]
\end{prop}

Sequential reduction is usually the standard notion of reduction studied in the literature on TRSs; indeed, one way to interpret $\to$ is as a simple model of computation and program evaluation. Notice that according to this reading, we obtain a formal definition of computation as $\to^*$, with 
$\to$ actually modeling single computation steps. There are, however, other possible notions of reduction one can consider. Here we recall two of the main ones akin to forms of \emph{parallel} and 
\emph{nested} computation.

\begin{deff}\label{def_parred}
Given a TRS $(\Sigma, V, \mapsto)$, \emph{parallel reduction} is the relation $\Rightarrow \,\,\subseteq \termset \times \termset$ inductively defined as: 
 \[ \dfrac{{\tt x} \in V}{\rel x \Rightarrow x}\quad \quad 
 \dfrac{\rel t \trianglearrow s}{\rel t \Rightarrow s}\quad\quad
 \dfrac{\rel {t_i}\Rightarrow {s_i} \quad\quad i \in \{1..n\}\quad\quad o \in \Sigma_n}{o({\tt t_1 \hdots t_n})\Rightarrow o({\tt s_1 \hdots s_n})}\]
\end{deff}

\begin{remark}
 The above definition defines $\Rightarrow$ inductively, without first going through a context-based definition as we did for $\to$. Such a definition is possible (and the reader can find it in any textbook on rewriting), however it requires several term-based notations that are highly bureaucratic and 
 syntax-dependent, and thus we have omitted it. A similar remark applies to 
 the definition of full reduction we are going to define.
\end{remark}

Note that the first defining clause in Definition \ref{def_parred} makes $\Rightarrow$ reflexive, whereas parallelism is actually given by the third clause. It is a routine exercise to show that every sequential reduction can be mimicked by a parallel one. A bit more challenging is to prove that a single parallel reduction can be mimicked by (usually) several sequential reductions. Altogether, these results state that the notion of computation introduced by parallel and sequential computation is one and the same.

\begin{prop}
 Given a TRS $(\Sigma, V, \mapsto)$, we have:
 \[ \to \,\, \subseteq\,\, \Rightarrow\,\, \subseteq \,\,\to^*\,\, = \,\,\Rightarrow^* \]
\end{prop}

Finally, we mention a further notion of reduction, called \emph{full reduction}.\footnote{The terminology 
is not uniform in the literature. For instance, full reduction is called \emph{parallel reduction} 
in the literature on higher-order reduction systems, whereas it is sometimes called 
\emph{multi-step} reduction in the setting of term rewriting. Here, we use the terminology 
of \cite{Gavazzo/LICS/2023}.}

\begin{deff}\label{def_fullred}
Given a TRS $(\Sigma, V, \mapsto)$, \emph{full reduction} is the relation $\Rrightarrow \,\,\subseteq \termset \times \termset$ inductively defined as:
 \[ \dfrac{{\tt x} \in V}{\rel x \Rrightarrow x}\quad \quad 
 \dfrac{\rel t \trianglearrow s}{\rel t \Rrightarrow s}\quad\quad
 \dfrac{\rel {t_i}\Rrightarrow {s_i} \quad o({\tt s_1 \hdots s_n}) \mathrel{\trianglearrow^{\mathit{refl}}} {\tt u}
 \quad\quad i \in \{1, \hdots, n\}\quad\quad o \in \Sigma_n}{o({\tt t_1 \hdots t_n})\Rrightarrow \tt u}\]
\end{deff}
Given a term of the form $o({\tt t_1}, \hdots, {\tt t_n})$, 
full reduction first evaluates all its subterms $\tt t_i$, say obtaining term $\tt s_i$, 
and then try to apply a ground reduction (instance) to the term 
$o({\tt s_1}, \hdots, {\tt s_n})$. Notice that this is not mandatory, as the latter term is 
reduced using $\trianglearrow^{\mathit{refl}}$, meaning that we can also stop the reduction 
process by reducing $o({\tt s_1}, \hdots, {\tt s_n})$ to itself. 

It is a straightforward exercise to prove that $\Rrightarrow$ is reflexive and, most importantly, 
subsumes parallel reduction. Additionally, as before,  
full and parallel (and sequential) reductions induce the same notion of computation.

\begin{prop}
\label{prop:spectrum-of-reduction-concrete}
 Given a TRS $(\Sigma, V, \mapsto)$, we have:
 \[ \to \,\, \subseteq\,\, \Rightarrow\,\, \subseteq\,\, \Rrightarrow \,\, \subseteq \,\,\to^*\,\, = \,\,\Rightarrow^*\,\, = \,\, \Rrightarrow^*\]
\end{prop}


\begin{example} The following are examples of the three kinds of reduction described. First, sequential reduction:
 \[ \tt M(\underline{M(0, 0)}, A(S(x), y)) \to M(\underline{0}, A(S(x), y))\]
Parallel reduction:
 \[ \tt M(\underline{M(0, 0)}, \underline{A(S(x), y)}) \Rightarrow M(\underline{0}, \underline{S(A(x, y))}) \]
Full reduction: 
 \[ \tt \underline{M(\underline{M(0, 0)}, {A(S(x), y)})} \Rrightarrow \underline{0}. \]
\end{example}

Now that we have recalled the necessary background notions on TRSs, we extend algebras of relations with suitable operators that will allow to define TRSs in a fully relation-algebraic fashion.

\section{Towards an Algebra of Term-Relations}
The goal of this section is to obtain a relational definition of TRS. Earlier we mentioned how algebras of relations are not expressive enough to describe contexts and substitutions. To study TRSs relationally, it is necessary to define an \emph{extended} algebra of relations called an \emph{algebra of term relations}. This algebra extends the traditional algebra of relations 
with operators whose intended semantics reflects 
the action of main syntactic notions -- such as term formation via 
operation symbols, substitution, contexts, $\hdots$ -- on relations between terms. 
The axioms of such operators then capture the structural properties 
induced by the intended semantics. 
Before proceeding any further, it is important to make explicit the methodology we 
shall follow in the remaining part of this manuscript (both in the present and 
subsequent chapters).

\begin{remark}[On Methodology]
Since the idea behind algebras of term relations is 
quite difficult to grasp, we follow a bottom-up approach, proceeding from the semantic model to 
the algebraic axiomatization. 
\begin{enumerate}
    \item 
    We first work with \emph{concrete} relations in $\Rel \termset$ and 
    extend the action of the main term-based operations -- viz. substitution and 
    term formation via operation symbols -- to \emph{relations} between terms. 
    This results in novel operators acting on relations, notably relation substitution 
    and compatible refinement. 
    \item 
    After that, we prove algebraic-like laws for such operators. Crucially, the validity of such laws 
    depends on core structural properties of their corresponding 
    term-based notions, such as syntax-preservation of 
    substitutions and alike. Most importantly, the algebraic-like laws fully capture 
    the aforementioned 
    structural properties, 
    in the sense that most proofs about rewriting can be given relying solely on those laws. 
    \item Finally, we abstract from the concrete case of relations on terms and define algebras of 
    term relations as traditional algebras of relations 
    extended with operators abstracting the ones in point $1$ 
    and with axioms along the laws in point $2$. This way, we obviously 
    recover $\Rel \termset$ as a concrete model of such algebras.    
    We then show how such algebras are expressive enough 
    to give algebraic and relational definitions of TRSs, of their notions of reductions, and to 
    prove properties about them in an equational style. 
\end{enumerate}
\end{remark}
 
\subsection{Substitution, Relationally}

We begin with relational substitution as first introduced by Lassen~\cite{Lassen/PhDThesis} in 
the context of $\lambda$-calculi. 
We first define it on concrete relations between terms and then distill its algebraic structure. 
 \begin{deff}  
 \emph{Relational substitution} is the map $-[\cdot] : \Rel \termset \times \Rel \termset \to \Rel \termset$ defined by:
 $$
 {\tt t} \mathrel{a[b]} {\tt s} \iff 
 \exists \sigma,\rho: V \to \termset. \exists 
 {\tt t}_0, {\tt s}_0.\ 
 \begin{cases}
    {\tt t} = {\tt t}_0^\sigma &
    \\
    {\tt s} = {\tt s}_0^\rho &
    \\
    {\tt t}_0 \mathrel{a} {\tt s}_0 &
    \\
    \forall {\tt x}.\ \sigma({\tt x}) \mathrel{b} \rho({\tt x}).
 \end{cases}
 $$
 In inference-style notation:
\[\dfrac{\rel {t}{a}{s}\quad\quad \forall {\tt x} \in V \, . \,\rel {\sigma (x)} b {\rho (x)}}{\rel {\subst \sigma t} {a[b]} {\subst \rho s}}\]
\end{deff}

\begin{remark}\label{rmk:subinst} 
Notice that, given a relation $a$, we obtain its substitution instances as 
$a[\Delta]$. In particular, $\trianglearrow$ as introduced in the previous section is precisely $\mapsto[\Delta]$. 
\end{remark}

  \begin{figure}
  \begin{framed}
  \begin{multicols}{2}
  \begin{equation}
  \Delta[\Delta] = 
  \Delta
  \end{equation}
  
\begin{equation}(a ; b) [a' ; b'] \leq a[a'] ; b[b']\end{equation}
  
\begin{equation}(a[b]) ^\circ = a^\circ [b^\circ]\end{equation}
  
\begin{equation}a \leq a', b \leq b' \implies a[b] \leq a'[b']\end{equation}
  
\begin{equation}\Big(\bigvee_i a_i\Big)[b] = \bigvee_i a_i[b]\end{equation}

\begin{equation} a[b][c] = a[b[c]] \end{equation}
  
  \end{multicols}
  
  \end{framed}
  \caption{Axioms of $-[\cdot]$ in algebras of Term-Relations}
  \label{fig:algebraic-laws-subst}
  \end{figure}

%
%
%
%
%
%
%
%
%
%

The algebraic laws of relation substitution have been first studied by 
Lassen~\cite{Lassen/PhDThesis} in the specific case of the $\lambda$-calculus 
and later generalized by Gavazzo~\cite{Gavazzo/LICS/2023} to general notions of 
syntax. We report such laws  in Figure \ref{fig:algebraic-laws-subst}.
 \begin{prop}[Algebra of relation substitution]
 The laws in Figure \ref{fig:algebraic-laws-subst} hold.
 \end{prop}

%

%
%
%
%
%
%
%
 
 \subsection{Compatibility, Relationally}

  A notion we need in order to study how ground reductions propagate along syntactic construct is that of \emph{compatible refinement}~\cite{Gordon/FOSSACS/01,Gordon-1995,Lassen/PhDThesis,Gavazzo/LICS/2023}. 
  This relates expressions that have the same outermost syntactic construct and pairwise related arguments. 
 \begin{deff} 
 \begin{itemize}
 \item[]
  \item  We define the map $\widetilde{\cdot}: \Rel \termset \to \Rel \termset$ by:
  $$
  {\tt t} \mathrel{\widetilde{a}} {\tt s} 
  \iff 
  \exists n \geq 0.\ \exists o \in \Sigma_n.\ \exists 
  {\tt t}_1, \hdots, {\tt t}_n, {\tt s}_1, \hdots, {\tt s}_n.\ 
  \begin{cases}
      {\tt t} = o({\tt t}_1, \hdots, {\tt t}_n) &
      \\
      {\tt s} = o({\tt s}_1, \hdots, {\tt s}_n) &
      \\
      \forall i \leq n.\ {\tt t}_i \mathrel{a} {\tt s}_i
  \end{cases}
  $$
  In inference-style:
  \[\dfrac{\rel {t_1} {a} {s_1} \hdots \rel {t_n} {a} {s_n}\quad o \in {\rm \Sigma^{n}} \quad n \geq 0}{o({\tt t_1, \hdots, t_n})\,\,\widetilde{ a}\,\, o({\tt s_1, \hdots, s_n})}\]  
  \item We define the relation $I_\eta \in \Rel \termset$ 
  as relating two terms if and only if they are the same variable:
  \[ \rel s { I_\eta}t \iff {\tt t} \in {V} \text{ and } \rel t = s \]
  In inference-style:
  \[\dfrac{{\tt x} \in V}{{\tt x} \mathrel{I_\eta} {\tt x}}\]  
 \end{itemize} 
 \end{deff}

The relation $\widetilde a$ relates only complex terms with the same outermost operation symbol
and pairwise $a$-related arguments. Notice that any relation of the form $\widetilde{a}$ 
relates any constant with itself.
Since terms are either variables or complex terms, we have also introduced the relation $I_\eta$. 
This relation is a \emph{coreflexive}~\cite{scedrov-freyd}, meaning that it is contained in the 
identity relation (i.e. $I_\eta \leq \Delta$). Accordingly, it can be regarded as the property of \emph{being a variable}. 
The notion of compatible refinement is obtained by joining $\widetilde{\cdot}$ and 
$I_\eta$.
 
 \begin{deff} Given $a \in \ralg$, we define its \emph{compatible refinement} $\widehat a$ as:
 \[\widehat a = I_\eta \vee \widetilde a\]
 That is:
 \[ \dfrac{{\tt x} \in V}{\rel x {\widehat{a}}  x} \quad\quad \dfrac{\rel {t_1} {a} {s_1} \hdots \rel {t_n} {a} {s_n}\quad o \in {\rm \Sigma^{ n}}}{o({\tt t_1, \hdots, t_n})\,\,\widehat{ a}\,\, o({\tt s_1, \hdots, s_n})} \]
 \end{deff}

  \begin{remark}
 The decomposition of $\widehat{-}$ into $\widetilde -$ and $I_\eta$ enhances the expressive power of relational reasoning. As an example, in the definition of terms, we have omitted the assumption that signatures and variables should be disjoint, which can be formalized as:
 \[ \widetilde a \wedge I_\eta = \bottom. \]
 Indeed, if there were terms such that $\rel t {(\widetilde a \wedge I_\eta)} s$, both $\rel t {\widetilde a} s$ and $\rel t {I_\eta} s$ would hold, i.e. $\tt t$ and $\tt s$ would be the same variable, say 
 $\tt x$, and a complex term at the same time. This is possible only if 
 $\tt x$ is an operation symbol of arity zero, hence showing that $\Sigma$ and $V$ 
 are not disjoint.
 Similarly, the decomposition allows us to express the property that a variable should never be a redex:
 \[ I_\eta ; \widetilde a = \bot \]
 \end{remark}
 
   
 
  
  
  

%
%
%
%
%
%

 \begin{figure}
\centering
\small
 \begin{framed}
 \begin{multicols}{2}
\begin{subfigure}[b]{0.4\textwidth}
      \begin{equation}\label{ax:widetildeDelta}
  \widetilde\Delta \leq \Delta
  \end{equation}
  \begin{equation}\label{ax:widetilde;}
  \widetilde {a;b} = \widetilde a ; \widetilde b
  \end{equation} 
  \begin{equation}\label{ax:widetildeop}
  \widetilde {a^\circ} = \widetilde a ^\circ
  \end{equation}
  \begin{equation}\label{ax:widetildemoncon}
  \text{$\widetilde -$ is monot. and $\omega$-cont.}
  \end{equation}
  \begin{equation}\label{ax:widetilde-[.]}
  \widetilde a [b] \leq \widetilde{a[b]}
  \end{equation}   
   \begin{equation}\label{ax:Ieta-tilde-bot}
 I_\eta \wedge \widetilde a = \bot
  \end{equation}
  \begin{equation}
      \widehat a = I_\eta \vee \widetilde a \label{ax:seqref-Ieta-tilde}
  \end{equation}
\end{subfigure}

\begin{subfigure}[b]{0.4\textwidth}
  \begin{equation}\label{ax:widehatDelta}
  \widehat\Delta = \Delta
  \end{equation}
  \begin{equation}\label{ax:widehat;}
  \widehat {a;b} = \widehat a ; \widehat b
  \end{equation}
  \begin{equation}\label{ax:widehatop}
  \widehat {a^\circ} = \widehat a ^\circ
  \end{equation}
  \begin{equation}\label{ax:widehatmoncon}
  \text{$\widehat -$ is monot. and $\omega$-cont.}
  \end{equation}
  \begin{equation}\label{ax:widehat-[.]}
  \widehat a [b] \leq \widehat{a[b]}\vee b
  \end{equation}
  \begin{equation}\label{ax:Ieta_corefl}
  I_\eta[b] \leq b
  \end{equation}
  \begin{equation}\label{ax:delta-fixpoint}
  \Delta = \mu x. \widehat x
  \end{equation}
\end{subfigure}
 \end{multicols}
 \end{framed}
   \caption{Axioms of $\widetilde -$ and $\widehat -$ in algebras of Term-Relations}
  \label{fig:algebraic-laws-compref}
 \end{figure}

Both $\widetilde -$ and $\widehat -$ satisfy several algebraic laws, which 
have been investigated in \cite{Gavazzo/LICS/2023}. 
 
\begin{prop}[Algebraic Laws]
The laws in Figure \ref{fig:algebraic-laws-compref} hold.
\end{prop}

Most of the rules in Figure \ref{fig:algebraic-laws-compref} are self-explanatory. There are, however, 
a couple of laws that deserve some comments. 
We begin with law \eqref{ax:widetilde-[.]} which expresses syntax-preservation of substitution. 
The latter can be expressed by the following equality:
\begin{equation}
    (o({\tt t}_1, \hdots, {\tt t}_n))^\sigma = 
    o({\tt t}_1^\sigma, \hdots, {\tt t}_n^\sigma)
    \label{syntax-preservation}
\end{equation}
It is not hard to see that the validity of \eqref{ax:widetilde-[.]} 
holds precisely in virtue of \eqref{syntax-preservation} and, moreover, that the former 
captures syntax-preservation of substitution at a relational level.

Another law that deserves a comment is law \eqref{ax:delta-fixpoint}. 
The latter has been proved by Lassen~\cite{Lassen/PhDThesis} and 
gives a relational counterpart of the binary induction principle by Jacobs \cite{Jacobs1997ATO}. 
In fact, \eqref{ax:delta-fixpoint} expresses the principle of structural recursion on terms, but 
at the level of relations. Recall the structural recursion on terms proof principle, where 
$A$ ranges over properties (i.e. subsets) of terms, $n \geq 0$, and $o \in \Sigma_n$:
\[ 
\dfrac{\forall {\tt x}. A({\tt x}) \qquad 
\forall {\tt t}_1, \hdots, {\tt t}_n.\ 
A({\tt t}_1) \mathrel{\&} \cdots \mathrel{\&} A({\tt t}_n) 
\implies A(o({\tt t_1, \hdots, t_n}))
}
{\forall {\tt t}.A({\tt t})} 
\]
Moving to relations, a property $A$ is given as a coreflexive $\alpha$ and proving that 
all terms have property $A$ means showing $\Delta \leq \alpha$. By \eqref{ax:delta-fixpoint}, 
we can prove $\Delta \leq \alpha$ using \eqref{KT-induction}, hence showing 
$\widehat{\alpha} \leq \alpha$, which amounts to prove 
$I_\eta \leq \alpha$ and $\widetilde{\alpha} \leq \alpha$. Unfolding the definitions of 
of $I_\eta$ and of the $\widetilde{-}$ operator, we see that the latter two inequalities 
are precisely 
the premises of the structural induction principle.


 \section {Algebras of Term Relations and Relational TRSs}
 We now have enough tools to reason about term rewriting in a purely relational and algebraic way. 
 We thus abstract from the concrete case of relations on terms (i.e. $\Rel \termset$) 
 and move to the abstract, axiomatic setting of algebras of relations. 
 To begin with, we extend the definition of (traditional) algebras of relation 
 with novel operators acting as abstract counterparts of the term-based operators previously introduced; 
 the defining axioms of such operators then capture all the relevant algebraic laws proved in the preceding sections.
 We refer to the algebraic structure thus obtained as \emph{algebra of term relations}~\cite{Gavazzo/LICS/2023}.
 
\begin{deff}[Algebra of Term Relations \cite{Gavazzo/LICS/2023}]
\label{aotr} 
An algebra of term relations is a structure 
\[\ralg = (A, \leq, ;, \Delta, -^\circ, I_\eta, \widetilde -, -[\cdot])\]
 such that:
 \begin{enumerate}[i.]
  \item $(A, \leq, ;, \Delta, -^\circ)$ is an algebra of relations.
  \item $I_\eta$ is a coreflexive, i.e. $I_\eta \leq \Delta$.
  \item $\widetilde-$ is a continuous unary operator on $\ralg$ satisfying:
  \begin{itemize}
   \item $\widetilde \Delta \leq \Delta$
   \item $\widetilde {a;b} = \widetilde {a}; \widetilde {b}$
   \item $\widetilde {a\op} = \widetilde a \op$
  \end{itemize}
 \item $-[\cdot]$ is a binary operator on $\ralg$, monotonic in both arguments and continuous in the first, and satisfying:
  \begin{itemize}
   \item $\Delta[\Delta] = \Delta$
   \item $a[b][c] = a[b[c]]$
   \item $(a ; b) [a' ; b'] \leq a[a'] ; b[b']$
   \item $(a[b]) ^\circ = a^\circ [b^\circ]$
  \end{itemize}
  \item The following laws hold, where $\widehat a$ abbreviates $I_\eta \vee \widetilde a$.
  \begin{multicols}{2}
  \begin{enumerate}[1.]
   \item \label{aotr_1} $\widetilde a[b] \leq \widetilde{a[b]}$
   \item \label{aotr_2}$I_\eta[b] \leq b$
   \item \label{aotr_3} $\Delta = \mu x. \widehat x$
   \item \label{aotr_4} $I_\eta \wedge \widetilde a = \bot$
  \end{enumerate}

  \end{multicols}
 \end{enumerate}
\end{deff}
The axioms in Definition~\ref{aotr} abstract the laws in Figure~\ref{fig:algebraic-laws-subst}
and Figure~\ref{fig:algebraic-laws-compref}, and thus their intended meaning should be clear. 

\begin{prop}
\label{prop:aotr-soundess-relset}
    The laws in Figure~\ref{fig:algebraic-laws-subst}
    and Figure~\ref{fig:algebraic-laws-compref} are derivable in any algebra of term relations.
\end{prop}
\begin{proof}
Most of the laws in Figure~\ref{fig:algebraic-laws-subst}
and Figure~\ref{fig:algebraic-laws-compref} correspond to defining axioms of an algebra of term relations, 
and thus are trivially derivable, whereas the remaining ones can be easily derived by simple 
calculations. 
\end{proof}

The intended meaning of the axioms in Definition~\ref{aotr} is the same as their corresponding concrete instances in 
$\Rel \termset$; and, in fact, Proposition~\ref{prop:aotr-soundess-relset} precisely tells us that 
$\Rel \termset$ (endowed with the relevant operations) gives a model of an algebra of term relations.\footnote{Further models
are given by higher-order~\cite{terese} (e.g. the $\lambda$-calculus~\cite{Barendregt/Book/1984}) 
and nominal~\cite{pitts-nominal-1} syntax, as proved in \cite{Gavazzo/LICS/2023}.}
It is nonetheless worth remarking that, contrary to concrete models, the axioms in Definition~\ref{aotr} 
are totally syntax-independent: they do not refer to any signature, set of variables or, more generally, to any syntactic notion.


\subsection{Term Rewriting Systems, Relationally}
Now the notion of algebra of term relations is defined, we can give a fully algebraic definition of a 
\emph{relational} TRSs.
 \begin{deff}[Relational TRS] Fixed an algebra of term-relations $\ralg$, an $\ralg$-TRS is given by a relation $a \in \ralg$ such that $I_\eta ; a = \bot$.
 \end{deff}
  
  \begin{remark}
 Note that $(\Sigma, V, \mapsto)$ is a TRS if and only if the relation $\mapsto \,\,\in \Rel \termset$ is an $\ralg$-TRS, because $\mapsto$ qualifying as a rewrite rule (i.e. a variable cannot be a redex) corresponds to $I_\eta;\mapsto \,\,= \bot$. This means that $\ralg$-TRSs model ground reductions.
  \end{remark}
  
  We now show that algebra of relations are expressive enough to model interesting notions of reductions, besides ground ones. First of all, recall from Remark \ref{rmk:subinst} that $\trianglearrow = \,\, {\mapsto}[\Delta]$, the substitution instances reduction can be modeled. Furthermore, by introducing the \emph{parallel closure}, we show that it is also possible to define parallel reduction. 

 \begin{deff}[Parallel closure]
 Fixed an algebra of Term-Relations $\ralg$, the parallel closure is:
 \[ \parclosure a\definedas \mu x. a \vee \widehat x \]
 \end{deff}

\begin{figure}
  \centering
\begin{framed}
\begin{multicols}{2}
     \begin{subfigure}[b]{0.3\textwidth}
         \centering
         \begin{align*}
            a &\leq \parclosure a\\
            \antihat {\parclosure a} &\leq {\parclosure a}\\
            \parclosure{(\parclosure a)} &= \parclosure a\\
            a \leq b &\implies \parclosure a \leq \parclosure b
        \end{align*}
         \caption{Closure laws}
         \label{fig:algebraic-laws-lattice-parclosure}
     \end{subfigure}

     \begin{subfigure}[b]{0.4\textwidth}
         \centering
            \begin{align*}
                \Delta &\leq \parclosure a\\
                \parclosure{(a ; b)} &\leq \parclosure a ; \parclosure b\\
                \parclosure a ; \parclosure b &\leq \parclosure{((a \vee \Delta); (b \vee \Delta))}
            \end{align*}         
        \caption{Monoid Laws}
         \label{fig:algebraic-laws-monoid-parclosure}
     \end{subfigure}
\end{multicols}
\end{framed}
        \caption{Derived Laws of Parallel Closure}
        \label{fig:algebraic-laws-parclosure}
\end{figure}

\begin{prop}[\cite{Gavazzo/LICS/2023}]
The laws in Figure \ref{fig:algebraic-laws-parclosure} hold.
\end{prop} 

\begin{deff}[Parallel Reduction]
Fixed an algebra of term relations $\ralg$ and a ground reduction $a \in \ralg$ (meaning that $I_\eta ; a = \bot$), 
we define the \emph{parallel reduction} induced by $a$ as $\parclosure{a[\Delta]}$.
 \end{deff}
 
  \begin{example}
 Note that, if $\ralg = \Rel \termset$, we have $\Rightarrow \,\, =  \parclosure\trianglearrow$ defined term-wise as:
  \[ \dfrac{\rel t \trianglearrow s}{\rel t{\parclosure \trianglearrow}s} 
  \quad\quad \dfrac{{\tt x} \in V}{\rel x {\parclosure \trianglearrow} x}
  \quad\quad \dfrac{\rel {t_1}{\parclosure \trianglearrow}{s_1} \text{ }\cdots\text{ } \rel{t_n}{\parclosure \trianglearrow}{s_n}\quad o \in \Sigma_n}{o({\tt t_1}, \hdots, {\tt t_n}) \,\,\parclosure \trianglearrow\,\, o({\tt s_1} \hdots {\tt s_1})}\]
 \end{example}

 A complete algebraic analysis of parallel reduction is beyond the scope of this manuscript: the interested reader 
 can consult \cite{Gavazzo/LICS/2023} for an extensive analysis of parallel reduction in the setting of algebra 
 of term relations. 

 Remarkably, algebras of term relations are also expressive enough to account for deep reductions. 

 \begin{deff}
    Given an algebra of term-relations $\ralg$,
     we define the unary operator $\fullclosure{\cdot}$ on it by
     $$
     \fullclosure a\definedas \mu x.\widehat{x}; a[\Delta]
     $$
 \end{deff}
 The $\mathsf{h}$-notation for the operator that ultimately defined full reduction comes 
 from a deep connection with the field of program equivalence: in the latter, the construction 
 $\fullclosure{\cdot}$ is known as \emph{Howe's method}~\cite{Howe/IC/1996,Pitts/ATBC/2011} and is
 a major operational technique to prove congruence properties of notions of program 
 equivalence. The connection between full reduction and Howe's method has been observed in
 \cite{Gavazzo/LICS/2023}, where Howe's technique has been used to prove confluence properties of 
 rewriting systems. We refer to that work for a deeper account of the operator  $\fullclosure{\cdot}$ 
 and its (nontrivial) algebraic laws. Here, we simply notice how we can use such an operator to 
 define full reduction in a purely algebraic fashion.

\begin{deff}[Full Reduction]
Fixed an algebra of term relations $\ralg$ and a ground reduction $a \in \ralg$ (meaning that $I_\eta ; a = \bot$), 
we define the \emph{full reduction} induced by $a$ as $\fullclosure{a[\Delta]}$.
 \end{deff}
 
 \begin{example}
 Note that, if $\ralg = \Rel \termset$, we have $\Rrightarrow \,\, =  \fullclosure\trianglearrow$ defined term-wise as:
  \[ \dfrac{\rel t \trianglearrow s}{\rel t{\fullclosure \trianglearrow}s} 
  \quad\quad \dfrac{{\tt x} \trianglearrow {\tt t}}{\rel {\tt x} {\fullclosure \trianglearrow} {\tt t}}
  \quad\quad \dfrac{\rel {t_1}{\fullclosure \trianglearrow}{s_1} 
  \text{ }\cdots\text{ } 
  \rel{t_n}{\fullclosure \trianglearrow}{s_n} \quad o \in \Sigma_n}
  {o({\tt t_1}, \hdots, {\tt t_n}) \,\,\fullclosure \trianglearrow\,\, o({\tt s_1} \hdots {\tt s_1})}
  \]
  \[
  \dfrac{\rel {t_1}{\fullclosure \trianglearrow}{s_1} 
  \text{ }\cdots\text{ } 
  \rel{t_n}{\fullclosure \trianglearrow}{s_n} \quad 
   o({\tt s_1} \hdots {\tt s_1}) \trianglearrow {\tt u}
  \quad o \in \Sigma_n}{o({\tt t_1}, \hdots, {\tt t_n}) \,\,\fullclosure \trianglearrow\,\, {\tt u}}
  \]
 \end{example}

 Notice that we could have avoided the second rule, as the constraint $I_\eta;a = \bot$ 
 (i.e. ${\tt t} \mapsto {\tt s}$ implies ${\tt t} \not \in V$, in the concrete case of 
 $\Rel \termset$) forbids it.

Finally, we observe that full reduction subsumes parallel reduction, but they both define the same notion 
of computation. This gives the first part of an abstract and algebraic account 
of Proposition~\ref{prop:spectrum-of-reduction-concrete}, whose complete statement and proof 
is a major contribution of this manuscript.

\begin{theo}[\cite{Gavazzo/LICS/2023}]
\label{theorem:spectrum-of-reduction-parallel-full}
    Let $\ralg$ be a sequential algebra of term relations. Then, we have the following 
    inequalities (spectra of reductions)
    \begin{align*}
        a \leq \parclosure a \leq \fullclosure a = {\parclosure a}^* = 
        {\fullclosure a}^*
        \\
        a \leq a[\Delta] \leq \parclosure{a[\Delta]} 
        \leq \fullclosure{a[\Delta]} = {\parclosure{a[\Delta]}}^* = 
        {\fullclosure{a[\Delta]}}^*
    \end{align*}
\end{theo}

\paragraph{Towards Sequential Reduction}
We have shown that algebras of term relations are capable of expressing parallel reduction, but from a computational point of view it is \emph{sequential} reduction that is more fundamental. In the next chapter we come to the very subject of this document, which is to define an algebra of term relations capable of dealing with sequential reduction.

\chapter{Sequential reduction}
\label{chapter:seq_red}

We now focus on sequential rule application, i.e. on those reduction that apply only one rule at a time. Our goal is to extend the algebra of term-relations defined in the previous chapter to obtain an algebra expressive enough to describe properties of sequential reductions. We do so by first introducing various operators on $\Rel \termset$ and proving their algebraic laws, and then relating the new operators to the ones previously defined, in order to finally be able to describe relationally the full spectrum of reductions.

\section{Sequential Refinement, Concretely}
 The notion that allows us to study the way sequential reduction relations propagate along syntactic construct is \emph{sequential refinement}. 
\begin{deff}[Sequential Refinement]\label{def:seqref} The function $\antihat {-} : \Rel \termset \to \Rel \termset$ is defined for all $a\in \Rel \termset$ and ${\tt{x}}, {\tt{y}} \in \termset$ as :
\begin{equation}
\rel {\repltbar}{\antihat a}{\replsbar} \iff \exists {\tt t}, {\tt s}, o.  \begin{cases}
                                             {\tt \repltbar} = o({\tt \halfvec{u_0}, t, \halfvec{u_1}})\\
                                             {\tt \replsbar} = o({\tt \halfvec{u_0}, s, \halfvec{u_1}})\\
o \in \Sigma_n : n \geq 1\\
\rel tas
                                            \end{cases} \label{seqrefII}
\end{equation}
i.e.: 
\begin{equation}
\label{seqref}
 \dfrac{\rel tas \quad\quad o \in \Sigma_n \quad n \geq 1}{o({\tt \halfvec{u_0}, t, \halfvec{u_1}})\,\,\antihat{a}\,\,o({\tt \halfvec{u_0}, s, \halfvec{u_1}})}
\end{equation}
\end{deff}

We now need to study the laws that relate the newly added operator to the usual operators of relation algebra, namely the identity $\Delta$, composition $;$, converse $-^\circ$, and how it behaves with respect to $\leq$ and $\vee$.
\newpage
\begin{prop}\label{antihat_id} 
$\antihat \Delta \leq \Delta$.
 \begin{proof} 
 \begin{align*}
\rel{\repltbar}{\antihat \Delta}{\replsbar} &\overset{\eqref{seqrefII}}\implies \exists {\tt t}, {\tt s}, o. \Big({\tt \repltbar} = o({\tt \halfvec{u_0}, t, \halfvec{u_1}}) \,\,\&\,\,{\tt \replsbar} = o({\tt \halfvec{u_0}, s, \halfvec{u_1}})\,\,\&\,\,\ o \in \Sigma_n\,\,\&\,\,{\rel t\Delta s}\Big)\\
&\overset{\eqref{identity_deff}}\implies
\exists {\tt t}, {\tt s}, o. \Big({\tt \repltbar} = o({\tt \halfvec{u_0}, t, \halfvec{u_1}}) \,\,\&\,\,{\tt \replsbar} = o({\tt \halfvec{u_0}, s, \halfvec{u_1}})\,\,\&\,\,\ o \in \Sigma_n\,\,\&\,\,{\tt{t} = \tt{s}}\Big)\\
&\implies
\exists {\tt t}, o. \Big({\tt \repltbar} = o({\tt \halfvec{u_0}, t, \halfvec{u_1}}) \,\,\&\,\, {\tt \replsbar} = o({\tt \halfvec{u_0}, t, \halfvec{u_1}})\,\,\&\,\,\ o \in \Sigma_n\Big)\\
&\implies {\tt \repltbar} = {\tt \replsbar} \\&\overset{\eqref{identity_deff}}\implies {\rel {\repltbar} \Delta {\replsbar}}
 \end{align*}
 \end{proof}
\end{prop}

 \begin{remark}
 The converse ($\Delta \leq \antihat\Delta$) is false, since $\antihat -$ only relates terms of non-null arity, so $\rel k {\antihat \Delta} k$ does not hold for any constant $\tt k$ (neither does it hold for variables).
 \end{remark}

\begin{prop}
\label{compI}
For all $a,b \in \Rel \termset$, $\antihat{ a ; b} \leq  {\antihat{a}} ; {\antihat{b}}$. 
 
 \begin{proof}\,
 
 $
   \rel {\repltbar} {\antihat { a ; b}} {\replsbar} \overset{\eqref{seqrefII}}\implies \exists {\tt t}, {\tt s}, o. \Big({\tt \repltbar} = o({\tt \halfvec{u_0}, t, \halfvec{u_1}}) \,\,\&\,\,{\tt \replsbar} = o({\tt \halfvec{u_0}, s, \halfvec{u_1}})\,\,\&\,\,\ o \in \Sigma_n\,\,\&\,\,{\rel t {( a;b)} s}\Big)
  $
  
 $\rel t {( a;b)} s \overset{\eqref{composition_deff}}\implies \exists {\tt r}. (\rel tar) \,\,\& \,\,(\rel rbs)$, therefore we may apply the definition of seq. ref.:
   
   \begin{multicols}{2}
   \[ \dfrac{\rel tar \quad\quad o \in \Sigma_n \quad n \geq 1}{\underbrace{o({\tt \halfvec{u_0}, t, \halfvec{u_1}})}_{\tt \repltbar}\,\,\antihat{a}\,\,o({\tt \halfvec{u_0}, r, \halfvec{u_1}})} \]
   
   \[ \dfrac{\rel rbs \quad\quad o \in \Sigma_n \quad n \geq 1}{o({\tt \halfvec{u_0}, r, \halfvec{u_1}})\,\,\antihat{b}\,\,\underbrace{o({\tt \halfvec{u_0}, s, \halfvec{u_1}})}_{\tt \replsbar}} \]
   \end{multicols}
   
   By \eqref{composition_deff}, we obtain $\rel {\repltbar} {({\antihat a} ; {\antihat b})} {\replsbar}$.
 \end{proof}
\end{prop}
\smallskip

\begin{remark}
 
The converse is false, consider for instance the system of arithmetic defined earlier, the reduction
\begin{align*}
 \tt M(\underbrace{\tt A(S0, 0)}_t, A(SS0, 0)) &\,\,\antihat \mapsto \,\,\tt M(S0, \underbrace{\tt A(SS0, 0)}_s) \tag {$\tt A(X, 0) \mapsto X$}\\
 &\,\,\antihat \mapsto \,\,\tt M(S0, SS0)\tag {$\tt A(X, 0) \mapsto X$}
\end{align*}
does not imply $\tt M({\tt A(S0, 0)}, A(SS0, 0)) \,\,\antihat{\mapsto ; \mapsto}\,\, M(S0, \tt A(SS0, 0))$. \bigskip

In general, assume $\rel {\repltbar}{({\antihat a} ; {\antihat b})} {\replsbar}$. Then by \eqref{composition_deff} there exists $\tt \replrbar$ such that $\rel {\repltbar}{\antihat a} {\replrbar} \,\, \&\,\,\rel {\replrbar}{{\antihat b}} {\replsbar}$.

 \[\rel {\repltbar}{\antihat a}{\replrbar} \iff \exists {\tt t}, {\tt r_0}.  \begin{cases}
                                             {\tt \repltbar} = o({\tt \halfvec{u_0}, t, \halfvec{u_1}})& o \in \Sigma_n\\
                                             {\tt \replrbar} = o({\tt \halfvec{u_0}, r_0, \halfvec{u_1}})\\
                                             \rel tar_0
                                            \end{cases} 
\]

\[\rel {\replrbar}{\antihat b}{\replsbar} \iff \exists {\tt r_1}, {\tt s}.  \begin{cases}
                                             {\tt \replrbar} = o({\tt \halfvec{v_0}, r_1, \halfvec{v_1}}) & o \in \Sigma_n\\
                                             {\tt \replsbar} = o({\tt \halfvec{v_0}, s, \halfvec{v_1}})\\
                                             \rel {r_1} bs
                                            \end{cases} 
\]
But there is no guarantee that:
\begin{itemize}
 \item $\tt r_0 = \tt r_1$
 \item $\tt r_0$ and $\tt r_1$ are at the same position in the operator
\end{itemize}
so we cannot conclude $\rel t {(a;b)} s$, which would imply $\rel {\repltbar} {(\antihat{ a;b})} {\replsbar}$.
\end{remark}

\begin{prop}
 \label{compII}
For all $a,b \in \Rel \termset$, $ {\antihat a}\, ; {\antihat b} \leq  {\antihat{ a ; b}} \vee { {\antihat b} ; {\antihat a}}$.
\end{prop}
\begin{proof} 

\[\rel {\repltbar} {\antihat a}{\replrbar} \,\, \& \,\, \rel {\replrbar} {\antihat b}{\replsbar}  \implies \exists \tt {t}, \tt{s}, \tt{r_0} , {\tt r_1}. \begin{cases}
                                             {\tt \repltbar} = o({\tt \halfvec{v_0}, t, \halfvec{v_1}}) & o \in \Sigma_n\\
                                             {\tt \replrbar} = o({\tt \halfvec{v_0}, r_0, \halfvec{v_1}}) \\
                                             {\tt \replrbar} = o({\tt \halfvec{u_0}, r_1, \halfvec{u_1}}) \\
                                             {\tt \replsbar} = o({\tt \halfvec{u_0}, s, \halfvec{u_1}})\\
                                             \rel ta{r_0}\\
                                             \rel {r_1} bs
                                            \end{cases} \]
The terms $\tt r_0$ and $\tt r_1$ defined above can either be \emph{in the same position} in the operator or not.

\begin{itemize}
 \item If $\tt r_0$ and $\tt r_1$ appear in the same position in $\tt z$, then: 
 \[\tt r_1 = r_0\quad \halfvec{v_0} = \halfvec{u_0}\quad \halfvec{v_1} = \halfvec{u_1}\]
 
Which by \eqref{composition_deff} implies $\rel t {(a; b)} s$, which in turn, by \eqref{seqref}, implies $\rel {\repltbar} {(\antihat{ a;b})} {\replsbar}$.
\item If $\tt r_0$ and $\tt r_1$ appear in different positions $i, j$ in $\tt z$, with $i \neq j$, we may represent the terms $\tt \repltbar$, $\tt \replsbar$, $\tt \replrbar$ as:
\[ {\tt \repltbar} = o(\hdots, \underset{i}{\tt t}, \hdots, \underset{j}{\tt r_1}, \hdots) \quad\quad {\tt \replsbar} = o(\hdots, \underset{i}{\tt r_0}, \hdots, \underset{j}{\tt s}, \hdots) \quad\quad {\tt \replrbar} = o(\hdots, \underset{i}{\tt r_0}, \hdots, \underset{j}{\tt r_1}, \hdots) \]
By definition of sequential refinement we then get:
   {\small \begin{multicols}{2}
   \[ \dfrac{\rel {r_1}{b}{s} \quad o \in \Sigma_n \quad n \geq 1}{\underbrace{o(\hdots, {\tt t}, \hdots, {\tt r_1}, \hdots)}_{\tt \repltbar}\,\,\antihat{b}\,\,o(\hdots, {\tt t}, \hdots, {\tt s}, \hdots)} \]
   
    \[ \dfrac{\rel {t}{a}{r_0} \quad o \in \Sigma_n \quad n \geq 1}{o(\hdots, {\tt t}, \hdots, {\tt s}, \hdots)\,\,\antihat{a}\,\,\underbrace{o(\hdots, {\tt r_0}, \hdots, {\tt s}, \hdots)}_{\tt \replsbar}} \]
   \end{multicols}}
Which, by \eqref{composition_deff} implies $\rel {\repltbar} {(\antihat b \,; \antihat a)} {\replsbar}$.
\end{itemize}
By combining these cases, we obtain $ {\antihat a}\, ; {\antihat b} \leq  {\antihat{ a ; b}} \vee { {\antihat b} ; {\antihat a}}$.
\end{proof}

\begin{prop} \label{antihat_converse}
For all $a \in \Rel \termset$, ${\antihat a\op} = \antihat{a\op}$.
\end{prop}

 \begin{proof}

   \[\begin{aligned}
   \rel {\repltbar} {\antihat {a}\op} {\replsbar} &\overset{\eqref{op_deff}}\iff \rel {\replsbar} {\antihat {a}} {\repltbar} \\ &\overset{\eqref{seqrefII}}\iff \exists {\tt t}, {\tt s}, o. \Big({\tt \repltbar} = o({\tt \halfvec{u_0}, t, \halfvec{u_1}}) \,\,\&\,\,{\tt \replsbar} = o({\tt \halfvec{u_0}, s, \halfvec{u_1}}) \,\,\&\,\, o \in \Sigma_n\,\,\&\,\,{\rel sat}\Big)
   \end{aligned}
  \]

  Since $\rel sat \iff \rel t{a\op}s$, we may apply the definition of sequential refinement (Definition \ref{def:seqref}) and obtain:

  \[\dfrac{\rel t{a\op}s \quad\quad o \in \Sigma_n \quad n \geq 1}{\rel {\repltbar} {\antihat{a\op}} {\replsbar}}\]
 \end{proof}

\begin{prop}
For all $a,b \in \Rel \termset$, $ a \leq b \implies  {\antihat a} \leq {\antihat b}$.
\end{prop}

 \begin{proof}
    \[
   \rel {\repltbar} {\antihat a} {\replsbar} \overset{\eqref{seqrefII}}\implies \exists {\tt t}, {\tt s}, o. \Big({\tt \repltbar} = o({\tt \halfvec{u_0}, t, \halfvec{u_1}}) \,\,\&\,\,{\tt \replsbar} = o({\tt \halfvec{u_0}, s, \halfvec{u_1}}) \,\,\&\,\, o \in \Sigma_n\,\,\&\,\,{\rel t {a} s}\Big)
  \]

  Since $a \leq b$, $\rel tas \implies \rel tbs$, which by \eqref{seqref} implies $\rel {\repltbar} {\antihat b} {\replsbar}$.
 \end{proof}

\begin{prop}\label{antihat_lax_pres}
For all $a_i \in \Rel \termset$,
$\bigvee_i (\antihat a_i) \leq \antihat {\bigvee_i  a_i}$.
\end{prop}

 \begin{proof} This follows from monotonicity, by Remark \ref{lax_pres_join}.
 \end{proof}

 \begin{prop}[Continuity] 
 For all $a_i \in \Rel \termset$,
 $\antihat{\bigvee_i{a_i}} = \bigvee_i\antihat{{a_i}}$.
 \end{prop}
 \begin{proof}
  One side of the equality holds by Proposition \ref{antihat_lax_pres}. We prove the other. 
  \[
   \rel {\repltbar} {\antihat {\bigvee_i{a_i}}} {\replsbar} \overset{\eqref{seqrefII}}\implies \exists {\tt t}, {\tt s}, o. \Big({\tt \repltbar} = o({\tt \halfvec{u_0}, t, \halfvec{u_1}}) \,\,\&\,\,{\tt \replsbar} = o({\tt \halfvec{u_0}, s, \halfvec{u_1}}) \,\,\&\,\, o \in \Sigma_n\,\,\&\,\,{\rel t {\bigvee_i{a_i}} s}\Big)
  \]
  Since $\rel t {(\bigvee_i{a_i})} s$, there exists an index $i$ such that $\rel t {a_i} s$. By \eqref{seqref}: $\rel {\repltbar} {\antihat a_i} {\replsbar} $, hence:
  
  \[\rel {\repltbar} {\bigvee_i \antihat{a_i}} {\replsbar}\]
 \end{proof}

\section{The theory of sequential algebras of Term-Relations}

\begin{figure}
\begin{framed}
\begin{multicols}{2}
 \begin{equation}\label{ax:antihatDelta} \antihat \Delta \leq \Delta\end{equation}
 
 \begin{equation}\label{ax:antihat;}\antihat{ a ; b} \leq  {\antihat{a}} ; {\antihat{b}}\end{equation}
 
 \begin{equation}\label{ax:;antihat} {\antihat a}\, ; {\antihat b} \leq  {\antihat{ a ; b}} \vee { {\antihat b} ; {\antihat a}}\end{equation}
 
 \begin{equation}\label{ax:antihatop}{\antihat a\op} = \antihat{a\op} \end{equation}

\begin{equation}\label{ax:antihatmon} a \leq b \implies  {\antihat a} \leq {\antihat b}\end{equation}

\begin{equation}\label{ax:antihatcon}\antihat{\bigvee_i{a_i}} = \bigvee_i\antihat{{a_i}}\end{equation}
\end{multicols}
\end{framed}
\caption{Axioms of $\antihat .$ in sequential algebras of Term-Relations}
\label{seqref::fig}
\end{figure}

%
%
%
%

Now that we have shown the properties of $\antihat .$ on $\Rel \termset$, we can extend the notion of algebra of Term-Relations. 

\begin{deff}\label{def:seqalgtr} A \emph{sequential algebra of Term-Relations} \[\ralg = (A, \leq, ;, \Delta, -^\circ, I_\eta, \widetilde -, -[.], \antihat .)\]
consists of an algebra of Term-Relations $(A, \leq, ;, \Delta, -^\circ, I_\eta, \widetilde -, -[.])$ and a binary operation $\antihat . \colon A \times A \to A$ such that the axioms in Figure \ref{seqref::fig} hold.
\end{deff}

The equivalent of parallel closure $\parclosure{(-)}$ in the context of sequential reductions is the similarly defined \emph{sequential closure}. We shall now study its algebraic properties, and the ways it interacts with the other constructs of our relation algebra. In the following, we fix an arbitrary sequential algebra of term-relations $\ralg = (A, \leq, ;, \Delta, -^\circ, I_\eta, \widetilde -, -[.], \antihat .)$. 

\begin{deff} For any $a \in A$, the function $F_a\colon A \to A$ is defined as \[F_a(x) \definedas a \vee \antihat x\]
\end{deff}

\begin{prop} For any $a \in A$, the function $F_a(\cdot)$ is monotonic and $\omega$-continuous. 
\label{as_existence}
\label{existence}
\end{prop}
\begin{proof} The function $F_a(x)$ can be decomposed as the join of $\antihat -$ with the $a$-constant operator on $A$. Since they are both monotonic, so is their join, and thus so is $F_a$ by Prop. \ref{join_of_monotonic}. By Prop. \ref{join_of_omegacont} one proves $\omega$-continuity similarly.
\end{proof}

By virtue of the above proposition, the least fixed point of $F_a$ exists and one can thus introduce the following.

\begin{deff} \label {seq-closure-def}
For any $a \in A$, its \emph{sequential closure}, in symbols $\seqclosure{a}$, is defined as 
\[\seqclosure{a} \definedas \mu F_a \text{ (i.e., }\mu x . a \vee \antihat x\text{)}\]
 \end{deff}


 The next proposition shows that $\seqclosure{(\cdot)}$ is a closure operator, so that $\seqclosure a$ gives the least sequentially compatible relation containing $a$, where a relation $a$ is sequentially compatible if $\antihat a \leq a$.
 
 \begin{prop} The following properties hold for all $c \in A$:
 \begin{enumerate}[i.]
 \label{properties}
 \begin{multicols}{2}
  \item \label{seqclos_prop1} $c \leq \seqclosure{c}$
  \item \label{seqclos_prop2}$\antihat {\seqclosure{c}} \leq \seqclosure{c}$ 
  \item \label{seqclos_prop3}$\seqclosure{(\seqclosure{c})} \leq \seqclosure{c}$
\item $a \leq b \implies \seqclosure{a} \leq \seqclosure{b}$
 \end{multicols}
\end{enumerate}
 
Also notice how, by Propositions \ref{properties}.\ref{seqclos_prop1} and \ref{properties}.\ref{seqclos_prop3}: $\seqclosure{(\seqclosure{c})} = \seqclosure{c}$
 \end{prop}

\begin{proof} 
\begin{enumerate}
 \item []
\item As $\seqclosure{c}$ is a fixed point of $F_c(x) = c \vee \antihat x$: $\seqclosure{c} = c \vee \antihat {\seqclosure{c}}$. Therefore $c \leq \seqclosure{c}$.
\item $\seqclosure{c} = c \vee \antihat {\seqclosure{c}}$. Therefore $\antihat {\seqclosure{c}} \leq \seqclosure{c}$
\item  Since $\seqclosure{(\seqclosure{c})} = \mu x.\seqclosure{c} \vee \antihat x$, by \eqref{KT-induction}:
\[ F(\seqclosure{c}) \leq \seqclosure{c} \implies \mu F_c \leq \seqclosure{c}  \]
Thus it is enough to prove $F_c(\seqclosure{c}) \leq \seqclosure{c}$. Since ${F_c}_c(\seqclosure{c}) = c \vee \antihat {\seqclosure{c}}$:

\begin{itemize}
 \item $c \leq \seqclosure{c}$ by Proposition \ref{properties}.\ref{seqclos_prop1}
 \item $\antihat {\seqclosure{c}} \leq \seqclosure{c}$ by Proposition \ref{properties}.\ref{seqclos_prop2}
\end{itemize}

Therefore $F(\seqclosure{c}) = c \vee \antihat {\seqclosure{c}} \leq \seqclosure{c}$.

\item Let $a \leq b$. By  \eqref{KT-induction} the following holds:

\[ F_a(\seqclosure{b}) \leq \seqclosure{b} \implies \mu F_a \leq \seqclosure{b} \]

Hence it is enough to show $F_a(\seqclosure{b}) \leq \seqclosure{b}$. $F_a(\seqclosure{b}) = a \vee \antihat {\seqclosure{b}}$.

\begin{itemize}
 \item $a \leq \seqclosure{b}$ by the assumption and Proposition \ref{properties}.\ref{seqclos_prop1} ($a \leq b \leq \seqclosure{b}$)
 \item $\antihat {\seqclosure{b}} \leq \seqclosure{b}$ by Proposition \ref{properties}.\ref{seqclos_prop2}.
\end{itemize}

\end{enumerate}
\end{proof}

Next we show how $\seqclosure{(\cdot)}$ interacts with relation composition and converse.

\begin{prop}[Converse] 
For all $a\in A$,
\[\so a = \os a\text{.}\]
\end{prop}
\begin{proof} In order to prove the equality, the following must be shown:
\begin{enumerate}
 \item $\so a \leq \os a$
 \item $\os a \leq \so a$
\end{enumerate}

Notice that (1) is actually implied by (2):

\begin{align*}
 \so a &= \ooso a \tag{\ref{op_involution}}\\
 &\leq \osoo a \tag{2}\\
 &= \os a\tag{\ref{op_involution}}
\end{align*}

(2) can be shown by \eqref{KT-induction}, since $\mu F_{a\op} = \os a$, where $F_{a\op}(x) = a\op \vee \antihat x$:

\[ F_{a\op}\big(\so a\big) \leq \so a \implies \mu F_{a\op} \leq \so a \]
Note that:
\begin{align*}
F_{a\op}(\so a) &= a\op \vee \antihat {{\seqclosure{a}}}\op\\
&= a\op \vee \antihat {\so a}\tag{\ref{ax:antihatop}}
\end{align*}

\begin{itemize}
 \item By the monotonicity of $(\_)\op$: \quad\quad$a \leq \seqclosure{a}\implies a\op \leq \so a$
 \item While by Proposition \ref{properties}.\ref{seqclos_prop2}: \quad\quad$\antihat {\so a} \leq \so a$
\end{itemize}
Therefore $F_{a\op}\big(\so a\big) = a\op \vee \antihat {\so a} \leq \so a$
\end{proof}

\begin{prop}[Composition] For all $a,b \in A$, the following hold.
\begin{enumerate}[i.]
 \item \label{comp_1}$\seqclosure{(a;b)} \leq \seqclosure{a} ; \seqclosure{b}$
 \item $\seqclosure{a};\seqclosure{b} \leq (a ; b) \vee (a ; \antihat {\seqclosure{b}}) \vee (\antihat {\seqclosure{a}} ; b) \vee (\antihat{ \seqclosure{a} ; \seqclosure{b}}) \vee (\antihat {\seqclosure{b}} ; \antihat {\seqclosure{a}})$
\end{enumerate}
\end{prop}

\begin{proof}
\begin{enumerate}[i.]
 \item []
 \item By \eqref{KT-induction}:
\[ F_{a ; b}(\seqclosure{a} ; \seqclosure{b}) \leq \seqclosure{a} ; \seqclosure{b} \implies \mu F_{a;b} \leq \seqclosure{a} ; \seqclosure{b} \]
It is therefore enough to prove: $F_{a ; b}(\seqclosure{a} ; \seqclosure{b}) \leq \seqclosure{a} ; \seqclosure{b}$.

\begin{itemize}
 \item Since $;$ is continuous in both arguments, it is also monotonic:
\begin{multicols}{2}

 \begin{itemize}
  \item $a \leq \seqclosure{a} \implies a ; b \leq \seqclosure{a};b$
  \item $b \leq \seqclosure{b} \implies \seqclosure{a}; b \leq \seqclosure{a}; \seqclosure{b}$
 \end{itemize}

\end{multicols}

Hence $a; b \leq \seqclosure{a}; \seqclosure{b}$.

\item By Proposition \ref{compII}:\quad $\antihat{\seqclosure{a}; \seqclosure{b}} \leq \antihat{\seqclosure{a}} ; \antihat{\seqclosure{b}}$. By the monotonicity of $;$ it follows that \[\antihat {\seqclosure{a}} ; \antihat {\seqclosure{b}} \leq \seqclosure{a} ; \seqclosure{b}.\]Hence \[\antihat{\seqclosure{a}; \seqclosure{b}} \leq \seqclosure{a} ; \seqclosure{b}.\]
\end{itemize}
By these two points, it follows $F_{a ; b}(\seqclosure{a} ; \seqclosure{b}) = (a ; b \vee \antihat{\seqclosure{a}; \seqclosure{b}}) \leq \seqclosure{a} ; \seqclosure{b}$.

\item $\seqclosure{a} = a \vee \antihat {\seqclosure{a}}$, \quad$\seqclosure{b} = b \vee \antihat {\seqclosure{b}}$.
\begin{align*}
\seqclosure{a} ; \seqclosure{b} &=  (a \vee \antihat {\seqclosure{a}}) ;  (b \vee \antihat {\seqclosure{b}})\\
&= a ; (b \vee \antihat {\seqclosure{b}}) \vee \antihat {\seqclosure{a}} ; (b \vee \antihat {\seqclosure{b}}) \tag{Figure \ref{fig:algenbraic-laws-quantale} (cont. of $;$)}\\&=
(a ; b) \vee (a ; \antihat {\seqclosure{b}}) \vee (\antihat {\seqclosure{a}} ; b) \vee (\antihat {\seqclosure{a}} ; \antihat {\seqclosure{b}})\tag{Figure \ref{fig:algenbraic-laws-quantale} 
 (cont. of $;$)}\\
&\leq
(a ; b) \vee (a ; \antihat {\seqclosure{b}}) \vee (\antihat {\seqclosure{a}} ; b) \vee (\antihat{ \seqclosure{a} ; \seqclosure{b}}) \vee (\antihat {\seqclosure{b}} ; \antihat {\seqclosure{a}}) \tag{\ref{ax:;antihat}}\\
\end{align*}

\end{enumerate}
\end{proof}\bigskip

\begin{remark}[Continuity, $\omega$-continuity and adjoint] Since the operator $\antihat -$ is continuous, it follows that it is also $\omega$-continuous. From its continuity it also follows that it is a left adjoint, by Lemma \ref{continuous_adjoint}. There exists in fact an operator $\antihatadj . : \ralg \to \ralg$ such that:
\[ \antihat a \leq b \iff a \leq \antihatadj b \]
\end{remark}\noindent
%
%
%
%
Additionally we prove a result showing how $\antihat -$ behaves with respect to the Kleene star.

\begin{prop}\label{kleenestar_antihat}
For all $a\in A$, $\antihat{a^{*}} \leq {\antihat {a}}^{*}$.
\end{prop}

\begin{proof}[Proof]
Proving $\antihat{a^{*}} \leq {\antihat {a}}^{*}$ is equivalent to proving $a^{*} \leq \antihatadj{{\antihat {a}}^{*}}$. We show this by \eqref{KT-induction}, using $a^{*} = \mu x. \Delta \vee a;x$.
\[ \Delta \vee a;\antihatadj{{\antihat {a}}^{*}} \leq\antihatadj{{\antihat {a}}^{*}} \implies a^* \leq \antihatadj{{\antihat {a}}^{*}} \]

\begin {itemize}
 \item $\Delta \leq \antihatadj{{\antihat {a}}^{*}}$ is equivalent to $\antihat \Delta \leq {\antihat {a}}^{*}$, which is true as $\antihat \Delta \leq \Delta \leq {\antihat {a}}^{*}$.
 \item $a;\antihatadj{{\antihat {a}}^{*}} \leq \antihatadj{{\antihat {a}}^{*}}$ is equivalent to:
 
 \[\longantihat{a;\antihatadj{{\antihat {a}}^{*}}} \leq {\antihat a}^{*}\]
 The following derivation concludes:
  
 \begin{align*}
  \longantihat{a;\antihatadj{{\antihat {a}}^{*}}} \leq {\antihat a}^{*} &\Longleftarrow \antihat a ; \longantihat{\antihatadj{{\antihat {a}}^{*}}} \leq {\antihat a}^{*}\tag {\eqref{ax:antihat;} and trans. of $\leq$}\\
  &\Longleftarrow \antihat a ; {\antihat {a}}^{*} \leq {\antihat {a}}^{*} \tag{Lemma \ref{cancellation_lemma}: cancellation}
 \end{align*}
 Since $\antihat a ; {\antihat {a}}^{*} \leq {\antihat {a}}^{*}$ holds by definition of the Kleene star.
\end {itemize}
\end{proof}

%
%
%
%
\noindent
Relying on the algebraic properties of $\seqclosure{(.)}$, we now show that sequential closure laxly distributes over Kleene star. This result will be crucial to relate parallel and sequential reduction.

\begin{prop}[Kleene star]
For all $a\in A$,
\[\seqclosure{(a^{*})} \leq (\seqclosure{a})^{*}\text{.}\]
\end{prop}
%
%
%
%
%
%

\begin{proof}
 By Proposition \ref{kleenestar_antihat}, for all $a$ we have $\antihat{a^{*}} \leq (\antihat {a})^{*}$. Hence we can apply Lemma \ref{BonksLemma}, which, given that $a^s = \mu x . a \vee \antihat x$, gives us $\seqclosure{(a^{*})} \leq (\seqclosure{a})^{*}$. 
 
\end{proof}

\begin{figure}
 \begin{framed}
 \begin{multicols}{3}
 \begin{subfigure}{0.3\textwidth}
 \begin{align*}
 c &\leq \seqclosure{c}\\
 \antihat {\seqclosure{c}} &\leq \seqclosure{c}\\
 \seqclosure{(\seqclosure{c})} &\leq \seqclosure{c}\\
 a \leq b &\implies \seqclosure{a} \leq \seqclosure{b}
 \end{align*}
\end{subfigure}
\begin{subfigure}{0.3\textwidth}
 \begin{align*}
 \so a &= \os a\\
 \seqclosure{(a;b)} &\leq \seqclosure{a} ; \seqclosure{b}\\
 \seqclosure{(a^{*})} &\leq (\seqclosure{a})^{*}
 \end{align*}
\end{subfigure}
\begin{subfigure}{0.3\textwidth}
 \begin{align*}
  \seqclosure{a};\seqclosure{b} \leq (a ; b) &\vee (a ; \antihat {\seqclosure{b}})\\ &\vee (\antihat {\seqclosure{a}} ; b)\\&\vee (\antihat{ \seqclosure{a} ; \seqclosure{b}}) \\&\vee (\antihat {\seqclosure{b}} ; \antihat {\seqclosure{a}})
 \end{align*}
 \end{subfigure}
 \end{multicols}
 
   \end{framed}

 \caption{Derived laws of $\seqclosure a \definedas \mu x . a \vee \antihat x$}
 \label{seqclosure::fig}
\end{figure}
\noindent
The properties of sequential closure proved so far are summarised in Figure \ref{seqclosure::fig}.

\chapter{A Differential Analysis of Rewriting}
\label{chapter:DAoR}
Having defined and studied the properties of sequential refinement and sequential closure, it is natural to ask ourselves how these new operators relate to the ones we defined for parallel reduction. In particular, for traditional TRSs we have the fundamental theorem (Proposition~\ref{prop:spectrum-of-reduction-concrete})
\[ \to \,\, \subseteq \,\, \Rightarrow \,\, \subseteq \,\, \to^* \,\, = \,\, \Rightarrow^* \]
relating sequential and parallel reduction. We want to prove the relational version of this:
\[\seqclosure a \leq \parclosure a \leq {\seqclosure a}^* = {\parclosure a}^*.\]
In order to prove this result we need a change in perspective. We shall study sequential rewriting from a point of view based on the theory of functor derivatives, as outlined in \cite{Gavazzo/LICS/2023}. We skip the categorical foundations and proceed in a purely algebraic fashion, viewing the operator $\antihat .$ as a degenerate case of a binary operator $\deriv -\cdot$, that allows us to establish a link with the compatible refinement operator. This perspective additionally suggests the introduction of a family of relational operators that establish an additional link between sequential and parallel operators, in the form of ``Taylor expansion''.

In Sections \ref{sec:derivativecon} and \ref{sec:Taylor}, we will introduce the operations of derivative and Taylor expansion on the algebra of Term-Relations $\Rel \termset$ and we will study their algebraic properties. Later, in Section \ref{sec:funda}, we will generalize these operations by introducing  \emph{differential} algebras of Term-Relations, and we will prove two fundamental theorems relating sequential and parallel reduction.

\section{Derivative, concretely}\label{sec:derivativecon}
The \emph{derivative} operator is a generalization of sequential refinement (Definition \ref{def:seqref}). 
 It is convenient to fix some preliminary notation.

\begin{notation}
 Given two vectors of terms $\tt \halfvec t = (t_1 \hdots t_n)$, $\tt \halfvec s = (s_1 \hdots s_n)$ and a relation $a \in \Rel \termset$, the relation $\compwise a$ relates $\tt \halfvec t$ and $\tt \halfvec s$ if and only if for all $i \in \{1 \hdots n\}$ $\rel {t_i} a {s_i}$.
\end{notation}

\begin{deff}\label{derivative_deff} 
 The function
$\deriv \cdot - : \Rel \termset \times \Rel \termset \to \Rel \termset$
 is defined for all relations $a,b \in \Rel \termset$ and terms ${\tt t},{\tt s} \in \termset$ as
 \[\rel { t} {\deriv a b} { s} \iff \exists n \geq 1 . \exists o \in \Sigma_n . \exists {\tt \halfvec t_1}, {\tt \halfvec t_2}, {\tt \halfvec s_1}, {\tt \halfvec s_2}, {\tt u}, {\tt v}.
 \begin{cases}
 {\tt  t} = o({\tt \halfvec t_1, u, \halfvec t_2})\\
 {\tt  s} = o({\tt \halfvec s_1, v, \halfvec s_2})\\
 \rel {u} b {v}\\
 \rel {\halfvec t_1} {\compwise a} {\halfvec s_1}\\
 \rel {\halfvec t_2} {\compwise a} {\halfvec s_2}
 \end{cases}
\]
i.e.,
 \[\dfrac{ \rel {u} b {v}\quad
 \rel {\halfvec t_1} {\compwise a} {\halfvec s_1}\quad
 \rel {\halfvec t_2} {\compwise a} {\halfvec s_2}\quad o \in \Sigma_n\quad n \geq 1}{o({\tt \halfvec t_1, u, \halfvec t_2}) 
 \,\,\deriv ab \,\,
 o({\tt \halfvec s_1, v, \halfvec s_2})}\]
\end{deff}

%
%
%
%
%
%
%

\begin{remark}\label{rmk:ntihatderiv}
 By comparing Definitions \ref{def:seqref} and \ref{derivative_deff}, it is immediate to see that $\antihat a = \deriv \Delta a$ for all $a \in \Rel \termset$.
\end{remark}

\begin{remark}
Notice that the arity of the operator $o$ in Definition \ref{derivative_deff} cannot be zero, thus the derivative of any relation shall only relate complex terms, not variables or constants. 

\end{remark}

As for sequential refinement, we are interested in some of the algebraic properties of this new operator.

\begin{prop} 
$\deriv \cdot -$ is monotonic in both arguments.
\end{prop}
\begin{proof}
 \begin{enumerate}
 \item []
  \item We first prove that for all $a, b, c \in \ralg$,  $a \leq b \implies \deriv a c \leq \deriv b c$. \smallskip
  
  Assume $a \leq b$ and let $\rel { t} {\deriv ac} { s}$. By Definition \ref{derivative_deff}, 
  
  there exist an $n$-ary operation $o$ and terms ${\tt \halfvec t_1}, {\tt \halfvec t_2}, {\tt \halfvec s_1}, {\tt \halfvec s_2}, {\tt u}, {\tt v}$ such that 
\[{\tt  t} = o({\tt \halfvec t_1}, {\tt u}, {\tt \halfvec t_2})\quad {\tt  s} = o({\tt \halfvec s_1}, {\tt v}, {\tt \halfvec s_2}) \quad \rel {\halfvec t_1} {\compwise a} {\halfvec s_1}\quad \rel u c v \quad \rel {\halfvec t_2} {\compwise a} {\halfvec s_2}\]
  
Since $a \leq b$, $\rel {\halfvec t_1} {\compwise a} {\halfvec s_1}$ implies $\rel {\halfvec t_1} {\compwise b} {\halfvec s_1}$, and $\rel {\halfvec t_2} {\compwise a} {\halfvec s_2}$ implies $\rel {\halfvec t_2} {\compwise b} {\halfvec s_2}$. By Def. \ref{derivative_deff}:

 \[\dfrac{ \rel {u} c {v}\quad
 \rel {\halfvec t_1} {\compwise b} {\halfvec s_1}\quad
 \rel {\halfvec t_2} {\compwise b} {\halfvec s_2}\quad o \in \Sigma_n\quad n \geq 1}{o({\tt \halfvec t_1, u, \halfvec t_2}) 
 \,\,\deriv bc \,\,
 o({\tt \halfvec s_1, v, \halfvec s_2})}\]

\item We now prove that for all $a, b, c \in \ralg$,  $a \leq b \implies \deriv c a \leq \deriv c b$. \smallskip

Assume $a \leq b$ and let $\rel { t} {\deriv ca} { s}$. Similarly to the previous step, by Def. \ref{derivative_deff}, there exist an $n$-ary operation $o$ and terms ${\tt \halfvec t_1}, {\tt \halfvec t_2}, {\tt \halfvec s_1}, {\tt \halfvec s_2}, {\tt u}, {\tt v}$ such that 
\[{\tt  t} = o({\tt \halfvec t_1}, {\tt u}, {\tt \halfvec t_2})\quad {\tt  s} = o({\tt \halfvec s_1}, {\tt v}, {\tt \halfvec s_2}) \quad \rel {\halfvec t_1} {\compwise c} {\halfvec s_1}\quad \rel u a v \quad \rel {\halfvec t_2} {\compwise c} {\halfvec s_2}\]
Since $a \leq b$, By Definition \ref{derivative_deff} we have:

 \[\dfrac{ \rel {u} b {v}\quad
 \rel {\halfvec t_1} {\compwise c} {\halfvec s_1}\quad
 \rel {\halfvec t_2} {\compwise c} {\halfvec s_2}\quad o \in \Sigma_n\quad n \geq 1}{o({\tt \halfvec t_1, u, \halfvec t_2}) 
 \,\,\deriv cb \,\,
 o({\tt \halfvec s_1, v, \halfvec s_2})}\]

 \end{enumerate}
 
\end{proof}

\begin{prop} 
$\deriv \Delta \Delta \leq \Delta$.
 \end{prop}

\begin{proof}
 By Remark \ref{rmk:ntihatderiv}, $\deriv \Delta \Delta = \antihat \Delta$. By Proposition \ref{antihat_id} we have $\antihat \Delta \leq \Delta$.
\end{proof}

\begin{prop} 
For all $a,a',b,b'\in \Rel \termset$,
$\deriv {(a ; a')} {b ; b'} \leq \deriv a b ; \deriv {a'}{b'}$.
\end{prop}

\begin{proof}
Assume $\rel { t} {\deriv {(a ; a')} {b ; b'}} { s}$. Then, by Definition \ref{derivative_deff}, there exist an $n$-ary operation $o$ and terms ${\tt \halfvec t_1}, {\tt \halfvec t_2}, {\tt \halfvec s_1}, {\tt \halfvec s_2}, {\tt u}, {\tt v}$ such that:
\[{\tt  t} = o({\tt \halfvec t_1}, {\tt u}, {\tt \halfvec t_2})\quad {\tt  s} = o({\tt \halfvec s_1}, {\tt v}, {\tt \halfvec s_2}) \quad \rel {\halfvec t_1} {\compwise {a;a'}} {\halfvec s_1}\quad \rel u {b;b'} v \quad \rel {\halfvec t_2} {\compwise {a;a'}} {\halfvec s_2}\]
By \eqref{composition_deff}, there exist terms $\tt \halfvec r_1$, $\tt \halfvec r_2$, $\tt w$ such that:
\[ \rel{\halfvec t_1} {\compwise a} {\halfvec r_1},\quad \rel{\halfvec r_1} {\compwise {a'}} {\halfvec s_1}, \quad \rel{\halfvec t_2} {\compwise a} {\halfvec r_2}, \quad \rel{\halfvec r_2} {\compwise {a'}} {\halfvec s_2},\quad \rel ubw, \quad \rel w {b'} v\]
Hence, by Definition \ref{derivative_deff}:
 \[\dfrac{ \rel {u} b {w}\quad
 \rel {\halfvec t_1} {\compwise a} {\halfvec r_1}\quad
 \rel {\halfvec t_2} {\compwise a} {\halfvec r_2}\quad o \in \Sigma_n\quad n \geq 1}{o({\tt \halfvec t_1, u, \halfvec t_2}) 
 \,\,\deriv ab \,\,
 o({\tt \halfvec r_1, w, \halfvec r_2})}\]
 
  \[\dfrac{ \rel {w} {b'} {v}\quad
 \rel {\halfvec r_1} {\compwise {a'}} {\halfvec s_1}\quad
 \rel {\halfvec r_2} {\compwise {a'}} {\halfvec s_2}\quad o \in \Sigma_n\quad n \geq 1}{o({\tt \halfvec r_1, w, \halfvec r_2}) 
 \,\,\deriv {a'}{b'} \,\,
 o({\tt \halfvec s_1, v, \halfvec s_2})}\]
By \eqref{composition_deff} we obtain $\rel t {\deriv a b ; \deriv {a'}{b'}} s$.
\end{proof}

\begin{prop}
For all $a,b\in \Rel \termset$, $\deriv a b \op = \deriv{a\op}{b\op}$.
\end{prop}
\begin{proof}
 Given any two terms $\tt t$, $\tt s$: 
 \[\rel t {\deriv a b \op} s \iff\rel s {\deriv a b} t  \]
 which holds if and only if there exist an operator $o \in \Sigma_n$, with $n\geq 1$, and terms $\tt \halfvec t_1,\halfvec t_2, \halfvec s_1,\halfvec s_2, u, v$ such that $\tt t = (\halfvec t_1, u ,\halfvec t_2)$, $\tt s = (\halfvec s_1, v ,\halfvec s_2)$ and:
 \[ \rel {\halfvec s_1}{\compwise a}{\halfvec t_1} \quad\quad \rel {\halfvec s_2}{\compwise a}{\halfvec t_2}\quad\quad \rel vbu \]
Which is equivalent to:
 \[ \rel {\halfvec t_1}{\compwise a\op}{\halfvec s_1} \quad\quad \rel {\halfvec t_2}{\compwise a\op}{\halfvec s_2}\quad\quad \rel u{b\op}v \]
 Which, by definition of $\deriv - \cdot$, holds if and only if $\rel t{\deriv{a\op}{b\op}}s$.
\end{proof}

The following theorem relates the derivative operator with the $\widetilde-$ operator.
 Before stating the theorem, we recall the definition of $\widetilde -$ on $\Rel \termset$:
\[\dfrac{\rel {t_1} {a} {s_1} \hdots \rel {t_n} {a} {s_n}\quad o \in {\rm \Sigma^{ n}}}{o({\tt t_1, \hdots, t_n})\,\,\widetilde{ a}\,\, o({\tt s_1, \hdots, s_n})}\] 
Additionally, we define the relation $I_{\Sigma_0}$ as the identity on constants:
 \[\dfrac{o \in \Sigma_0}{o \,\,I_{\Sigma_0} \,o}\]

\begin{theo} \label{deriv_tilde} 
For all $a\in \Rel \termset$,
$\widetilde a = \deriv a a \vee I_{\Sigma_0}$.
\end{theo}
\begin{proof} By dividing the definition of $\widetilde -$ into the cases $n = 0$, $n > 0$, we obtain the following definition:
 \[ \dfrac{o \in \Sigma_0}{o \,\,{\widetilde{a}_0}\,  o} \quad\quad \dfrac{\rel {t_1} {a} {s_1} \hdots \rel {t_n} {a} {s_n}\quad o \in {\rm \Sigma^{ n}} \quad n > 0}{o({\tt t_1, \hdots, t_n})\,\,\widetilde{a}_{1}\,\, o({\tt s_1, \hdots, s_n})} \]
 And $\widetilde a = \widetilde a_0 \vee \widetilde a_{1}$. It is immediate that $\widetilde a_0 = I_{\Sigma_0}$, and $\widetilde a_1 = \deriv aa$. Hence $\widetilde a = \deriv a a \vee I_{\Sigma_0}$.
\end{proof}


The algebraic laws of $\deriv{\cdot}{-}$ discovered so far are summarised in Figure \ref{deriv::fig}. 


\begin{figure}
\begin{framed}
\begin{multicols}{2}
\centering \small
\begin{subfigure}{0.5 \textwidth}
 \begin{equation}\label{ax:deriv-monot}
 \deriv \cdot - \text{ is monot. in both arguments}
 \end{equation}
 \begin{equation}\label{ax:derivDelta}
 \deriv \Delta \Delta \leq \Delta 
 \end{equation}
 \begin{equation} \label{ax:deriv;}
 \deriv {(a ; a')} {b ; b'} \leq \deriv a b ; \deriv {a'}{b'}
 \end{equation}
  \begin{equation} \label{ax:derivop}
 \deriv a b \op = \deriv{a\op}{b\op}
 \end{equation}
 \end{subfigure}
 \begin{subfigure}{0.5 \textwidth}
  \begin{equation}\label{ax:;deriv}
  \deriv \Delta a ; \deriv \Delta b \leq \deriv \Delta {a;b} \vee \deriv \Delta b ; \deriv \Delta a
  \end{equation}
  \begin{equation}\label{ax:ftc}
\widetilde a = \deriv a a \vee I_{\Sigma_0}
  \end{equation}
\begin{equation}\label{ax:derivdelta-cont}
    \derivdelta{-} \text{ is continuous}
\end{equation}
 \end{subfigure}
\end{multicols}
\end{framed}
\caption{Axioms of $ \deriv {\cdot} {-}$ and $I_{\Sigma_0}$ in differential algebras of Term-Relations}
\label{deriv::fig}
\end{figure}

\section{Taylor Expansion, concretely}\label{sec:Taylor} 

We now introduce a further differential-like relational operator, akin to Taylor expansion \cite{trench2013introduction}. Later in this chapter, such operator shall be used to link our theory of sequential reduction with Gavazzo's theory of parallel and full reduction.

\begin{deff} [Taylor Expansion] \label{taylor_def} 

For all natural numbers $n \in \mathbb N$, the function $\tay n - : \Rel \termset \to \Rel \termset$ is defined as follows:
\[\dfrac{\rel {t_1} {a} {s_1} \hdots \rel {t_n} {a} {s_n}\quad o \in {\rm \Sigma^{ n}}}{o({\tt t_1, \hdots, t_n})\,\,{ \tay na }\,\, o({\tt s_1, \hdots, s_n})}\]
\end{deff}
At first this definition may look the same as that of $\widetilde a$, but notice that, in this case, the relation depends on $n$. In particular $\widetilde a$ relates terms the outermost operator of which could be of any arity, whereas with $\tay n a$ the arity must always be $n$.


Like for sequential refinement and derivatives, we investigate hereafter the algebraic properties of $\tay n -$. 

\begin{prop}
$\tay n \Delta \leq \Delta$
\end{prop}
\begin{proof}
  Assume $\rel t {\tay n \Delta} s$. Then there exist an $n$-ary operator $o$ and terms $\tt t_1 \hdots t_n$, $\tt s_1 \hdots s_n$ such that:
  \[{\tt t} = o({\tt t_1 \hdots t_n})\quad {\tt s} = o({\tt s_1 \hdots s_n})\quad \rel {t_i} \Delta {s_i}\] for all $i$, i.e. $\tt t_i = s_i$. Hence ${\tt t} = o({\tt t_1 \hdots t_n}) = o({\tt s_1 \hdots s_n}) = {\tt s}$, that is, $\rel t \Delta s$.
\end{proof}

\begin{prop}
For all $a,b\in \Rel \termset$, $\tay n {a;b} = \tay n {a} ; \tay nb$.
\end{prop}
\begin{proof}
 Assume $\rel t {\tay n {a;b}} s$. Then there exist an $n$-ary operator $o$ and terms $\tt t_1 \hdots t_n$, $\tt s_1 \hdots s_n$ such that:
 \[{\tt t} = o({\tt t_1 \hdots t_n})\quad {\tt s} = o({\tt s_1 \hdots s_n})\quad \rel {t_i} {a;b} {s_i}\] 
 for all $i$. This implies that there exist terms $\tt r_1 \hdots r_n$ such that for all $i$ $\rel {t_i} a {r_i}$ and $\rel {r_i} a {s_i}$. Therefore, by Definition \ref{taylor_def}:
 \[ {\tt t} = o({\tt t_1 \hdots t_n}) \,\,
 \tay n a \,\, o({\tt r_1 \hdots r_n})\quad\quad
 o({\tt r_1 \hdots r_n}) \,\,
 \tay n b \,\, o({\tt s_1 \hdots s_n}) = {\tt s}
\]
Which, by \eqref{composition_deff}, implies $\rel t {\tay n {a} ; \tay nb} s$. 

We now prove the other inclusion. Assume $\rel t {\tay n {a} ; \tay nb} s$. Then there exists $\tt r$ such that $\rel t {\tay n {a}} r$ and $\rel r {\tay n {b}} s$. This, by Definition \ref{taylor_def}, implies that there exist an $n$-ary operator $o$ such that:
 \[{\tt t} = o({\tt t_1 \hdots t_n})\quad {\tt r} = o({\tt r_1 \hdots r_n}) \quad {\tt s} = o({\tt s_1 \hdots s_n})\]
 and, for all $i$, $\rel {t_i} a {r_i}$ and $\rel {r_i} a {s_i}$. By \eqref{composition_deff}, we obtain: $\rel {t_i} {(a;b)} {s_i}$, which by Definition \ref{taylor_def} implies $\rel t {\tay n {a;b}} s$.
\end{proof}

\begin{prop}
\label{tay_conv_I}
For all $a\in \Rel \termset$, $\tay n {a\op} \leq {\tay n a}\op$.
\end{prop}
\begin{proof}
 Assume $\rel t {\tay n {a\op}} s$. Then by Definition \ref{taylor_def} there exist an $n$-ary operator $o$ and terms $\tt t_1 \hdots t_n$, $\tt s_1 \hdots s_n$ such that:
 \[{\tt t} = o({\tt t_1 \hdots t_n})\quad {\tt s} = o({\tt s_1 \hdots s_n})\quad \rel {s_i} a {t_i} \text{, for all $i$.}\] 
 Hence, by Definition \ref{taylor_def}, $\rel s {\tay na} t$, i.e. $\rel t {(\tay na)\op} s$
\end{proof}

\begin{prop}
For all $a\in \Rel \termset$, ${\tay n a}\op \leq \tay n {a\op}$.
\end{prop}

\begin{proof}
 Let $\rel t {{\tay n a}\op} s$. 
 \begin{align*}
  \rel t {{\tay n a}\op} s &\implies \rel s {\tay n a} t \tag{\ref{op_deff}}\\
  &\implies \rel s {\tay n {{a\op}\op}} t\tag{\ref{op_involution}}\\
  &\implies \rel s {(\tay n {a\op})\op} t\tag{Proposition \ref{tay_conv_I}}\\
  &\implies \rel t {(\tay n {a\op})} s\tag{\ref{op_deff}}\\
 \end{align*}
\end{proof}

\begin{prop} $\tay n -$ is monotonic: For all $a,b\in \Rel \termset$,
\[a \leq b \implies \tay n a \leq \tay n b\text{.}\]
\end{prop}
\begin{proof}
 Let $\rel t {\tay n a} s$ and $a \leq b$. By Definition \ref{taylor_def}, there exist an $n$-ary operator $o$ and terms $\tt t_1 \hdots t_n$, $\tt s_1 \hdots s_n$ such that:
 \[{\tt t} = o({\tt t_1 \hdots t_n})\quad {\tt s} = o({\tt s_1 \hdots s_n})\quad \rel {s_i} a {t_i} \text{, for all $i$.}\] 
 As $a \leq b$, this implies that, for all $i$, $\rel {s_i} b {t_i}$, which by Definition \ref{taylor_def} implies $\rel t {\tay n b}s$.
\end{proof}

\begin{prop} 
For all $a\in \Rel \termset$, $\tay 0 a = I_{\Sigma_0}$. 
\end{prop}
\begin{proof} This follows immediately from the definition of $\tay 0 a$, which is precisely the same as that of $I_{\Sigma_0}$:
\[\dfrac{o \in \Sigma_0}{\rel {\it o} {\tay 0 a} {\it o}}\]
\end{proof}

Recall the definition of $a[b]$:
 \[\dfrac{\rel {t}{a}{s}\quad\quad \rel {v_1} b {w_1} \hdots \rel {v_n} b {w_n}}{{\tt t[v_1 \hdots v_n / x_1 \hdots x_n]} \,\,{ a}[{b}]\,\,  {\tt s[w_1 \hdots w_n / x_1 \hdots x_n]}}\]

\begin{prop}
For all $a,b\in \Rel \termset$, $(\taynpar n a) [b] \leq \tay n {a[b]}$.
\end{prop}
\begin{proof} Let $\sigma, \rho$ be two substitutions defined as:
\[\tt \subst \sigma t = t[v_1 \hdots v_n / x_1 \hdots x_n] \quad \tt \subst \rho t = t[w_1 \hdots w_n / x_1 \hdots x_n]\]
Assume $t = o ({\tt t_1 \hdots t_n})$, $s = o ({\tt s_1 \hdots s_n})$ and $\rel{t^\sigma}{\big(\taynpar n a\big)[b]} {s^\rho}$. By the definitions of $\tay n -$ and $-[\cdot]$:
\[
\myinfer{\rel {t_1} a {s_1} \hdots \rel {t_n} a {s_n} \quad\quad \rel {\halfvec v}{\bf b}{\halfvec w} \quad\quad o \in \Sigma_n}{\subst \sigma {o ({\tt t_1 \hdots t_n})} \,\, (\taynpar n {a})[b] \,\, \subst \rho {o ({\tt s_1 \hdots s_n})}}
\]
And, by the same definitions, we conclude with the following:
\[
\myinfer{\myinfer{\rel {t_i} a {s_i} \quad \rel {\halfvec v}{\bf b}{\halfvec w}}{\rel{\subst \sigma {t_i}} {a[b]} {\subst \rho {s_i}}} \quad\quad i \in \{1 \hdots n\} \quad\quad o \in \Sigma_n
}{\subst \sigma {o ({\tt t_1 \hdots t_n})} \,\, \tay n {a[b]} \,\,\subst \rho{o ({\tt s_1 \hdots s_n})}}
\]
\end{proof}

It remains to study how the operator $\tay n -$ is related to sequential and compatible refinements. Let us begin with the former. Recall that the relation $\deriv \Delta a^{(n)}$ is defined as:
\[ \deriv \Delta a^{(n)} := \underbrace{\deriv \Delta a ; \hdots ; \deriv \Delta a}_n \]
i.e. $\deriv \Delta a^{(n)}$ relates terms $o({\tt \halfvec t})$ and $o({\tt \halfvec s})$ such that there exist $n$ non necessarily distinct components of $\tt\halfvec t$ and $\tt\halfvec s$, denoted $\{{\tt t_i}\}_{i \in I}$ and $\{{\tt s_i}\}_{i \in I}$, respectively, such that $\rel {t_i} a {s_i}$. 

\begin{prop}\label{taylor_deriv}
For all $a\in \Rel \termset$, $\tay n a \leq \deriv \Delta a^{(n)}$.
\end{prop}

\begin{proof}
For $n=0$, we have $\tay 0 a = I_{\Sigma_0} \leq \Delta \leq \deriv \Delta a^{(0)}$. For $n > 0$, let $o({\tt \halfvec t}) \,\,\tay n a\,\, o({\tt \halfvec s})$. This implies $\rel {t_i} a {s_i}$, for all $i$. Hence:
\begin{align*}
 o({\tt t_1 \hdots t_n}) \,\,\deriv \Delta a\,\, &o({\tt s_1 \,\, t_2 \hdots t_n})\\
 \deriv \Delta a\,\, &o({\tt s_1 \,\, s_2 \,\, t_3 \hdots t_n})\\
 \hdots\,\,\,\, &o({\tt s_1 \hdots s_n}),
\end{align*}
i.e. $o({\tt \halfvec t}) \,\,\deriv \Delta a ^{(n)}\,\, o({\tt \halfvec s})$.
\end{proof}

Finally, we can relate $\tay n -$ with $\widetilde -$ (and thus with $\widehat -$) relying on a Taylor expansion-like result.
\begin{theo}[Taylor Expansion] \label{taylor_tilde}
For all $a\in \Rel \termset$, $\widetilde a = \bigvee_n \tay n a$.
\end{theo}

\begin{proof} As noted earlier, the difference between the definitions of $\widetilde a$ and $\tay n a$ is that $\tay n a$ relates all terms related by $\widetilde a$ the outermost operator of which is $n$-ary. This is because their definitions are:
\[\dfrac{\rel {t_1} {a} {s_1} \hdots \rel {t_n} {a} {s_n}\quad o \in {\rm \Sigma^{ n}}}{o({\tt t_1, \hdots, t_n})\,\,\widetilde a\,\, o({\tt s_1, \hdots, s_n})} \quad\quad\quad\quad \dfrac{\rel {t_1} {a} {s_1} \hdots \rel {t_n} {a} {s_n}\quad o \in {\rm \Sigma^{ n}}}{o({\tt t_1, \hdots, t_n})\,\,{ \tay na }\,\, o({\tt s_1, \hdots, s_n})} \]
It is thus evident that, for all $n$, $\tay n a \leq \widetilde a$, which implies $ \bigvee_n \tay n a \leq \widetilde a$. For the other direction, it is enough to notice that for any two terms $\tt t, s$ such that $\rel t {\widetilde a} s$, their (common) outermost operator has finite arity $n$. Therefore $\rel t {\tay n a} s$. This means that for all relations $a$ there always exists $n$ such that $\widetilde a \leq \deriv n a$, which implies $\widetilde a \leq \bigvee_n \deriv n a$

\end{proof}

The algebraic laws of $\tay n -$ discovered so far are summarised in Figure \ref{tay::fig}.

\begin{figure}
\begin{framed}
\hspace{.4cm}
\begin{multicols}{2}
 \begin{equation}\label{ax:tayDelta}
 \tay n \Delta \leq \Delta
 \end{equation}
 
 \begin{equation} \label{ax:tay;}
 \tay n {a;b} = \tay n {a} ; \tay nb
 \end{equation}
 
  \begin{equation} \label{ax:tayop}
{\tay n a}\op = \tay n {a\op} 
\end{equation}

\begin{equation}\label{ax:taymon}
 a \leq b \implies \tay n a \leq \tay n b
\end{equation}

\begin{equation}\label{ax:tay0}
\tay 0 a = I_{\Sigma_0}
\end{equation}

\begin{equation}\label{ax:taysub}
(\taynpar n a) [b] \leq \tay n {a[b]}
\end{equation}

  \begin{equation}\label{ax:tayderiv}
\tay n a \leq \deriv \Delta a^{(n)}
  \end{equation}

  \begin{equation}\label{ax:tayexpansion}
\widetilde a = \bigvee_n \tay n a
  \end{equation}
\end{multicols}
\end{framed}
\caption{Axioms of $\tay n -$ in differential algebras of Term-Relations}
\label{tay::fig}
\end{figure}

\section{The theory of differential algebra of Term Relations}\label{sec:funda}

In Section \ref{sec:derivativecon}, we have illustrated that in the sequential algebra of Term-Relations $\Rel \termset$ both $\widetilde -$ and $\antihat -$ can be expressed by means of the derivative $\deriv -\cdot$ and the coreflexive $I_{\Sigma_0}$. In Section \ref{sec:Taylor}, we have introduced Taylor expansion $\tay n -$ and shown how this operation is related to $\widetilde -$. 

Hereafter, we extend the notion of algebra of Term-Relations by adding $\deriv -\cdot$, $I_{\Sigma_0}$ and $\tay n -$ and requiring that they satisfy the properties that we have proved so far.


%

\begin{deff} A \emph{differential algebra of Term-Relations} \[\ralg = (A, \leq, ;, \Delta, -^\circ, I_\eta, \widetilde -, -[.], I_{\Sigma_0}, \deriv -\cdot, \{ \tay n -\}_{n\in \omega})\]
consists of 
\begin{itemize}
\item an algebra of Term-Relations $(A, \leq, ;, \Delta, -^\circ, I_\eta, \widetilde -, -[.])$, 
\item a binary operation $\deriv -\cdot \colon A \times A \to A$ and a coreflexive $I_{\Sigma_0}\subseteq \Delta$ satisfying the axioms in Figure \ref{deriv::fig} and, 
\item for each $n \in \omega$, a unary operation $\tay n - \colon A \to A$ satisfying the axioms in Figure \ref{tay::fig}.
\end{itemize}
\end{deff}

Observe that in the definition above, we build upon an algebra of Term-Relations (Definition \ref{aotr}) rather than on a sequential one (Definition \ref{def:seqalgtr}). The operation $\antihat -$ can however be derived by fixing $\antihat a \definedas \deriv \Delta a$. Since all the axioms in Figure \ref{seqref::fig} can easily be derived by those in Figure \ref{deriv::fig}.


\begin{prop}
The laws in Figure \ref{seqref::fig} hold in any differential algebra of Term-Relations.
\end{prop}
\begin{proof}
From $\antihat x = \derivdelta x$ the following hold:
\begin{multicols}{2}
\begin{enumerate}[i.]
    \item \eqref{ax:antihatDelta} is equivalent to \eqref{ax:derivDelta},
    \item \eqref{ax:antihat;} follows from \eqref{ax:deriv;},
    \item \eqref{ax:;antihat} is equivalent to \eqref{ax:;deriv},
    \item \eqref{ax:antihatop} follows from \eqref{ax:derivop},
    \item \eqref{ax:antihatmon} follows from \eqref{ax:deriv-monot},
    \item \eqref{ax:antihatcon} is equivalent to \eqref{ax:derivdelta-cont}.
\end{enumerate}
\end{multicols}
\end{proof}
%
%
%


The following theorems, deemed \emph{fundamental}, link sequential and parallel reduction. In particular, they state that a sequential reduction can be performed as parallel reduction, and that any parallel reduction can be simulated through a finite amount of sequential steps. 

\begin{theo}[First Fundamental Theorem]\label{first_fun_theo}
 $\seqclosure a \leq \parclosure a$
\end{theo}
\begin{proof} By fixed point induction:
\[ a \vee \deriv \Delta {\parclosure a} \leq \parclosure a \implies \seqclosure a \leq \parclosure a \]
Since $a \leq a \vee \widehat{\parclosure a} = \parclosure a$, it is enough to show $\deriv \Delta {\parclosure a} \leq \parclosure a$.  
We have:
\begin{align*}
 \deriv \Delta {\parclosure a} &\leq \deriv  {\parclosure a} {\parclosure a} \tag{monot.  of $\deriv\cdot -$ \eqref{ax:deriv-monot} and $\Delta \leq \parclosure a$ (Figure \ref{fig:algebraic-laws-monoid-parclosure})}\\
 &\leq I_{\Sigma_0} \vee \deriv  {\parclosure a} {\parclosure a}\\
 &\leq \widetilde{\parclosure a} \tag{\ref{ax:ftc}}\\
 &\leq I_{\eta} \vee \widetilde{\parclosure a}\\
 &= \widehat {\parclosure a} \tag{\ref{ax:seqref-Ieta-tilde}}\\
 &\leq \parclosure a \tag {closure properties of $\parclosure -$ (Figure \ref{fig:algebraic-laws-monoid-parclosure})}
\end{align*}
\end{proof}

%
%
%
%
%

By Theorem \ref{first_fun_theo}, we know that in any differential algebra of term-relations we have $\seqclosure a \leq \parclosure a$. We now show $\parclosure a \leq {\seqclosure a}^{*}$.

\begin{theo}[Second Fundamental Theorem] \label{theorem:second_fund_theo}
 $\parclosure a \leq {\seqclosure a}^{*}$
\end{theo}
\begin{proof} 
By \eqref{KT-induction}:
\[ a \vee \widehat {{\seqclosure a}^{*}} \leq {\seqclosure a}^{*} \implies \parclosure a \leq {\seqclosure a}^{*} \]
As $a \leq \seqclosure a \leq {\seqclosure a}^{*}$, we only need to prove $\widehat{{\seqclosure a}^{*}} \leq {\seqclosure a}^{*}$, i.e. \[I_{\eta} \vee \widetilde{{\seqclosure a}^{*}} \leq {\seqclosure a}^{*}\]
Since $I_\eta \leq \Delta \leq {\seqclosure a}^{*}$, it is enough to show $\widetilde{{\seqclosure a}^{*}} \leq {\seqclosure a}^{*}$, which, by $\eqref{ax:tayexpansion}$, is equivalent to:
\[ \bigvee_n \tay n {{\seqclosure a}^{*}} \leq a^{\sf s*}\]
i.e. $\tay n {{\seqclosure a}^{*}} \leq {\seqclosure a}^{*}$, for all $n \geq 0$. We conclude by the following derivation:
\begin{align*}
 \tay n {{\seqclosure a}^{*}} &\leq \deriv \Delta {{\seqclosure a}^{*}}^{(n)}\tag{\ref{ax:tayderiv}}\\
 &\leq \deriv \Delta {{\seqclosure a}^{*}}^{*} \tag{\ref{a^n-in-astar}}\\
 &\leq {\deriv \Delta {\seqclosure a}}^{**} \tag {\ref{kleenestar_antihat}}\\
 &\leq {\deriv \Delta {\seqclosure a}}^{*} \tag{\ref{kleene-idemp}}\\
 &\leq {\seqclosure a}^{*} \tag{\ref{properties}.\ref{seqclos_prop2}, \ref{kleene-monot}}
\end{align*}

\end{proof}

\begin{corollary} ${\seqclosure a}^* = {\parclosure a}^*$.
\end{corollary}
\begin{proof} We know that $\seqclosure a \leq \parclosure a \leq {\seqclosure a}^{*}$. By monotonicity of the Kleene star, this implies ${\seqclosure a}^* \leq {\parclosure a}^* \leq {\seqclosure a}^{**}$. But ${\seqclosure a}^* = {\seqclosure a}^{**}$, hence ${\seqclosure a}^* = {\parclosure a}^*$.
 
\end{proof}

\paragraph{Spectrum of reductions} 
Finally, we can relate all the reductions introduced so far in the unifying framework of 
sequential algebras of term relations, this way obtaining an abstract and algebraic formulation 
of Proposition~\ref{prop:spectrum-of-reduction-concrete}.

\begin{theo}
\label{theorem:spectrum-of-reduction}
    Let $\ralg$ be a sequential algebra of term relations. Then, we have the following 
    inequalities (spectra of reductions)
    \begin{align*}
        a \leq \seqclosure a \leq \parclosure a \leq \fullclosure a =  {\seqclosure a}^* = {\parclosure a}^* = 
        {\fullclosure a}^*
        \\
        a \leq a[\Delta] \leq \seqclosure{a[\Delta]} \leq \parclosure{a[\Delta]} 
        \leq \fullclosure{a[\Delta]} =  {\seqclosure{a[\Delta]}}^* = {\parclosure{a[\Delta]}}^* = 
        {\fullclosure{a[\Delta]}}^*
    \end{align*}
\end{theo}

\begin{proof}
    By Theorem \ref{theorem:spectrum-of-reduction-parallel-full}, we have the following:
        \begin{align*}
        a \leq \parclosure a \leq \fullclosure a = {\parclosure a}^* = 
        {\fullclosure a}^*
        \\
        a \leq a[\Delta] \leq \parclosure{a[\Delta]} 
        \leq \fullclosure{a[\Delta]} = {\parclosure{a[\Delta]}}^* = 
        {\fullclosure{a[\Delta]}}^*
    \end{align*}
    By Theorems \ref{first_fun_theo} and \ref{theorem:second_fund_theo} we can conclude:
        \begin{align*}
        a \leq \seqclosure a \leq \parclosure a \leq \fullclosure a =  {\seqclosure a}^* = {\parclosure a}^* = 
        {\fullclosure a}^*
        \\
        a \leq a[\Delta] \leq \seqclosure{a[\Delta]} \leq \parclosure{a[\Delta]} 
        \leq \fullclosure{a[\Delta]} =  {\seqclosure{a[\Delta]}}^* = {\parclosure{a[\Delta]}}^* = 
        {\fullclosure{a[\Delta]}}^*
    \end{align*}
\end{proof}

\chapter{Sequential Reduction at Work: a Critical Pair-like Lemma}
In the previous chapters we described a novel theory of rewriting, introducing several operators that model different term-specific notions. This chapter serves as an example of how the theory of Differential Algebras of Term-Relations can be put into practice, which we chose to do by proving a lemma akin to the Critical Pair Lemma \cite{Huet80}. Before introducing such result, we prove an auxiliary lemma, which serves as a proof method for showing weak (i.e. local) confluence.
\label{chapter:CP-like}
 \begin{lemma}
 \label{lemma:proof-technique-weak-confluence}
 For all $a\in A$, if 
  \begin{multicols}{2}
\begin{equation}\label{eq:hyplemma1}\tag{$\dagger$} a\op ; a \leq {\seqclosure a} ^* ; {\so a} ^* \end{equation}

\begin{equation}\label{eq:hyplemma2}\tag{$\ddagger$} a\op ; \deriv \Delta{\seqclosure a} \leq {\seqclosure a} ^* ; {\so a}^*\end{equation}
 \end{multicols}\noindent
then $\so a ; \seqclosure a\leq {\seqclosure a} ^* ; {\so a} ^*$. 
 \end{lemma}

\begin{proof}
Observe that $\so a \overset{\text{Figure }\ref{seqclosure::fig}}= \os a = \mu x . a\op \vee \deriv \Delta x$.
Since composition is continuous (Figure \ref{fig:algenbraic-laws-quantale}), one can use enhanced $\omega$-continuous fixed point induction (Lemma \ref{ehanced-kleene-ind}, take $F(x)= a\op \vee \deriv \Delta x$ and $G(x) = x; \seqclosure a$). In this way,
in order to show $ \so a ; \seqclosure a \leq {\seqclosure a} ^* ; {\so a} ^*$, it is enough to prove
  \[ x \leq \so a \,\, \& \,\, x; \seqclosure a \leq {\seqclosure a} ^* ; {\so a} ^* \implies (a\op \vee \deriv \Delta x) ; \seqclosure a \leq {\seqclosure a} ^* ; {\so a} ^* \]
  for all $x\in A$. 
We thus assume $x \leq \so a $ and $ x; \seqclosure a \leq {\seqclosure a} ^* ; {\so a} ^*$, and prove:

  \begin{equation}
 \label{cp_1} a\op ; \seqclosure a  \, \leq \, {\seqclosure a} ^* ; {\so a} ^*
 \end{equation}
\begin{equation}
  \label{cp_2} \deriv \Delta x ; \seqclosure a \, \leq \, {\seqclosure a} ^* ; {\so a} ^*
\end{equation}
For \eqref{cp_1}, we have the following derivation.
\begin{align*}
a\op ; \seqclosure a &= a \op ; (a \vee \deriv \Delta {\seqclosure a}) \tag{Definition \ref{seq-closure-def}} \\
& = a\op ; a \vee a\op ; \deriv \Delta {\seqclosure a} \tag {\ref{composition_cont}} \\
& \leq ({\seqclosure a} ^* ; {\so a} ^*) \vee a\op ; \deriv \Delta {\seqclosure a} \tag {\ref{eq:hyplemma1}} \\
& \leq ({\seqclosure a} ^* ; {\so a} ^*) \vee ({\seqclosure a} ^* ; {\so a} ^*) \tag {\ref{eq:hyplemma2}}\\
&= {\seqclosure a} ^* ; {\so a} ^*
\end{align*}
For \eqref{cp_2}, observe that:
\begin{align*}
 \derivdelta x ; \seqclosure a &= \derivdelta x ; (a \vee \derivdelta {\seqclosure a}) \tag{Definition \ref{seq-closure-def}} \\
 &= \derivdelta x ; a \vee \derivdelta x ; \derivdelta {\seqclosure a} \tag {\ref{composition_cont}}
\end{align*}
Thus, it is enough to prove:
\begin{equation}
 \label{cp_a} \derivdelta x ; a \, \leq \, {\seqclosure a} ^* ; {\so a} ^*
 \end{equation}
\begin{equation}
  \label{cp_b} \derivdelta x ; \derivdelta {\seqclosure a} \, \leq \, {\seqclosure a} ^* ; {\so a} ^*
\end{equation}
For \eqref{cp_a}, we have the following derivation.
\begin{align*}
    \derivdelta x ; a &= ((\derivdelta x ; a)\op ) \op \tag{\ref{op_involution}} \\
    & = (( a\op ; \derivdelta x \op ) \op \tag{\ref{op_composition_rule}} \\
    & \leq ((  a \op ; \derivdelta {{\seqclosure a}\op }\op ) \op \tag{Hyp $x\leq {\seqclosure a}\op$}\\ \tag{\ref{ax:antihatop}}
& = (( { a} \op ; \derivdelta  {{\seqclosure a} } ) \op \\
    & \leq ({\seqclosure a}^* ; {\so a}^*)\op \tag{\ref{eq:hyplemma2}, \eqref{op_rule_deff}} \\
    &= {{\so a}^{*}}\op ; {{\seqclosure a}^*}\op \tag{\ref{op_composition_rule}}\\
&= {\soo a}^{*} ; {{\seqclosure a}^*}\op \tag{\ref{kleene-op}}\\
&= {{\seqclosure a}^*} ; {{\seqclosure a}^*}\op \tag{\ref{op_involution}}\\
&= {\seqclosure a}^* ; {\so a}^* \tag{\ref{kleene-op}}\\
\end{align*}
For \eqref{cp_b}, we have the following derivation.
\begin{align*}
\derivdelta x ; \derivdelta{\seqclosure a} & \leq \derivdelta {x ; \seqclosure a} \, \vee \, ( \derivdelta{\seqclosure a} ; \derivdelta x) \tag{axiom in Figure \ref{ax:;antihat}}\\
& \leq \derivdelta {({\seqclosure a}^* ; {\so a}^*)} \, \vee \, ( \derivdelta{\seqclosure a}  ; \derivdelta x ) \tag{Ind. Hyp: $x ; \seqclosure a \leq {\seqclosure a}^* ; {\so a}^*$} \\
&\leq \derivdelta {{\seqclosure a} ^*} ; \derivdelta {{\so a} ^*} \, \vee \, ( \derivdelta{\seqclosure a}  ; \derivdelta x ) \tag{\ref{ax:antihat;}, \ref{composition_cont}}\\
& = \derivdelta {{\seqclosure a} ^*} ; \derivdelta {{\os a} ^*} \, \vee \, ( \derivdelta{\seqclosure a}  ; \derivdelta x )\tag{$\os a = \so a$ (Figure \ref{seqclosure::fig}), \ref{composition_cont}}\\
 &\leq \derivdelta {{\seqclosure a}}^* ; \derivdelta {{\os a}}^*  \, \vee \, ( \derivdelta{\seqclosure a}  ; \derivdelta x ) \tag{\eqref{kleenestar_antihat}, \ref{composition_cont}}\\
 &\leq {\seqclosure a}^* ; {\os a}^* \, \vee \, ( \derivdelta{\seqclosure a}  ; \derivdelta x ) \tag {Definition \ref{seq-closure-def}}\\
 &= {\seqclosure a}^* ; {\so a}^* \, \vee \, ( \derivdelta{\seqclosure a}  ; \derivdelta x ) \tag{$\so a = \os a$ (Figure \ref{seqclosure::fig})} \\
 &\leq {\seqclosure a}^* ; {\so a}^* \, \vee \, ( \derivdelta {\seqclosure a} ; \derivdelta {\so a } )\tag{Ind. Hyp: $x \leq \so a$ }\\
 &\leq {\seqclosure a}^* ; {\so a}^* \, \vee \, ( \derivdelta {\seqclosure a} ; \derivdelta {\os a} )\tag{$\so a = \os a$}\\
 &\leq {\seqclosure a}^* ; {\so a}^* \, \vee \, ( \seqclosure a ; \os a) \tag{Definition \ref{seq-closure-def}}\\
 &\leq {\seqclosure a}^* ; {\so a}^* \, \vee \, ({\seqclosure a}^* ; {\so a}^*) \tag{\ref{kleene-a-in-astar}}\\
 & = {\seqclosure a}^* ; {\so a}^*
\end{align*}

\end{proof}

Now that we have a proof technique for local confluence, we can isolate conditions on TRSs 
that entail the hypothesis of Lemma~\ref{lemma:proof-technique-weak-confluence}. 
A milestone result on weak confluence for term rewriting system is the celebrated Critical Pair 
Lemma by Huet~\cite{Huet80}. Such a result states that if in a TRS all critical pairs are weakly confluent, then 
so is the whole system. The notion of a critical pair, unfortunately, is heavily syntactic and 
built upon specific details of first-order syntax\footnote{Different kinds of syntax require different notions 
of a critical pair.}, rather than on structural properties. 
For these reasons, we shall not work with critical pairs: instead, we are going to isolate some behavioural properties 
that are satisfied by any TRSs with weakly confluent critical pairs. We will then show how such conditions 
imply the 
hypothesis of Lemma~\ref{lemma:proof-technique-weak-confluence}, this way obtaining a general critical pair-like result. 

For ease of exposition, it is convenient to introduce a more compact notation for 
$a[\Delta]$.

\begin{notation}
    In the following, we write $a^{\mathsf{i}}$ for $a[\Delta]$.
\end{notation}

\begin{deff}
    Fixed a differential algebra of term-relations $\ralg$, we say that $\relone$ has 
    the critical pair property (CP) if:
    \begin{align}
        a^{\mathsf{i}\circ}; a^{\mathsf{i}} &\leq a^{\mathsf{is}*};a^{\mathsf{is}*\circ} 
        \label{cp-base}
        \tag{CP-1}
        \\
        a^{\mathsf{i}\circ}; \derivdelta{a^{\mathsf{is}}} &\leq \Delta[a^{\mathsf{is}}];a^{\mathsf{ih}\circ} 
        \label{cp-step}
        \tag{CP-2}
    \end{align}
\end{deff}

The reader familiar with the notion of critical pair will recognize that, when instantiated on $\Rel \termset$, 
law \eqref{cp-base} simply states 
that if a term ${\tt t}$ acts as a redex for two \emph{ground reduction instances}, so that 
${\tt t} \trianglearrow {\tt s}_1$ and ${\tt t} \trianglearrow {\tt s}_2$, then ${\tt s}_1$ and 
${\tt s}_2$ must have a common $\to^*$-reduct. 
Law \eqref{cp-step}, instead, considers the case in which we have a ground redex of the form 
${\tt t} = o({\tt t}_1, \hdots, {\tt t}_n)$ with a subterm ${\tt t}_i$ that can be $\to$-reduced. 
The notion of a critical pair is usually introduced to give a fine analysis of this kind of interference. 
Accordingly, there are two situations that can happen. In the first one, the interference is only apparent, 
as the nested redex that determines the reduction of ${\tt t}_i$ does \emph{de facto} come from 
a substitution of a variable in ${\tt t}$. In this case, one simply keeps track of that variable during 
the (ground) reduction of ${\tt t}$, and then reduce all its (possibly many) instances. 
It is also not hard to see that to do so, one can use a single step of full reduction rather than 
iteration of sequential reduction, exactly as prescribed by condition \eqref{cp-step}. 
In the second case, the interaction is real and precisely defines a so-called critical pair. In such a case, 
there is no way to overcome the problem, so that one can only impose (weak) convergence of critical pairs  
as a necessary hypothesis. Condition \eqref{cp-step} does so by asking critical pairs to converge 
by using full reduction but, of course, one can relax this requirement. 
The main point of  \eqref{cp-step} is to precisely capture how to achieve convergence in case 
of apparent overlaps of reductions. 

Notice also that using variations of \eqref{cp-base} and \eqref{cp-step}, one can define different classes of 
\emph{weakly confluent} TRSs. As an example, a TRSs without CPs is characterized precisely by law
\eqref{cp-step} and the following variation of law \eqref{cp-base}. 
\begin{align}
        a^{\mathsf{i}\circ}; a^{\mathsf{i}} &\leq \Delta
        \label{cp-base-prime}
        \tag{CP-1'}
    \end{align}
It is a straightforward exercise to prove that Proposition~\ref{prop:CP-lemma} below holds for TRSs 
satisfying \eqref{cp-step} and \eqref{cp-base-prime}, hence giving weak confluence of TRSs without 
critical pairs.

Let us now come back to our critical pair-like result. To prove it, we first recall a useful property of 
parallel reduction.
\begin{lemma}[\cite{Gavazzo/LICS/2023}]
\label{lemma:aux-cp-1}
$\Delta[a^{\mathsf{ip}}] \leq a^{\mathsf{ip}}$.
\end{lemma}

Finally, we show that that CP implies weak confluence.

\begin{prop}
\label{prop:CP-lemma}
    The critical pair property implies weak confluence.
\end{prop}
\begin{proof}
    The proof proceeds by instantiating Lemma~\ref{lemma:proof-technique-weak-confluence}
    to $a^{\mathsf{i}}$, hence obtaining the implication:
    $$
     a^{\mathsf{i}\circ}; a^{\mathsf{i}} \leq a^{\mathsf{is}*};a^{\mathsf{is}*\circ} 
     \text{ \& }
     a^{\mathsf{i}\circ}; \derivdelta{a^{\mathsf{is}}} \leq a^{\mathsf{is}*};a^{\mathsf{is}*\circ}  
     \implies 
    a^{\mathsf{is}\circ};a^{\mathsf{is}}  \leq a^{\mathsf{is}*};a^{\mathsf{is}*\circ} 
     $$
     Consequently, it is sufficient to prove
     \begin{align} 
     a^{\mathsf{i}\circ}; a^{\mathsf{i}} &\leq a^{\mathsf{is}*};a^{\mathsf{is}*\circ} 
     \label{cp-lemma-1}
     \tag{$\dag$}
     \\
     a^{\mathsf{i}\circ}; \derivdelta{a^{\mathsf{is}}} &\leq a^{\mathsf{is}*};a^{\mathsf{is}*\circ}
     \label{cp-lemma-2}
     \tag{$\ddag$}
     \end{align}
     Moreover, since \eqref{cp-lemma-1} is exactly \eqref{cp-base}, it is enough to prove 
     \eqref{cp-lemma-2}. We calculate:
     \begin{align*}
         a^{\mathsf{i}\circ}; \derivdelta{a^{\mathsf{is}}} \leq a^{\mathsf{is}*};a^{\mathsf{is}*\circ}
         &\impliedby \Delta[a^{\mathsf{is}}];a^{\mathsf{ih}\circ} \leq a^{\mathsf{is}*};a^{\mathsf{is}*\circ}
         \tag{\ref{cp-step}}
         \\
         &\impliedby \Delta[a^{\mathsf{ip}}];a^{\mathsf{ih}\circ} \leq a^{\mathsf{is}*};a^{\mathsf{is}*\circ}
         \tag{\ref{first_fun_theo} on $a^{\mathsf{i}}$ \& monotonicity of $-[\cdot]$}
         \\
         &\impliedby a^{\mathsf{ip}};a^{\mathsf{ih}\circ} \leq a^{\mathsf{is}*};a^{\mathsf{is}*\circ}
         \tag{\ref{lemma:aux-cp-1}}
         \\
         &\impliedby \text{Theorem \ref{theorem:spectrum-of-reduction-parallel-full}}
     \end{align*}
\end{proof}


\chapter{Conclusion}
For a long time, relational techniques have been applied to abstract rewriting. Together with \cite{Gavazzo/LICS/2023}, this manuscript shows that it is possible to use the same kind of techniques to reason about term-based rewriting through the introduction of \emph{syntax-inspired} operators. In spite of the intended semantics, the theory obtained is a purely algebraic and syntax-independent one. Syntax-independence is desirable because it makes for a very flexible theory; we, in fact, conjecture that the theory we developed also models notions of syntax with variable binding (e.g. the $\lambda$-calculus \cite{pierce2002types,Barendregt/Book/1984}) and the typed syntax of modern functional programming languages. Another advantage is the reduced logical complexity of proofs, most of which are expressed in a quasi-equational logic based on the substitution of equals for equals and as moderate a use of induction as possible. For this reason, we believe that it is possible to formalize our results by means of proof assistants, relying on libraries for relational reasoning \cite{pous}. 

\paragraph{Future Work} Much of the relational theory of TRSs is still to be developed, in particular in this manuscript we focused on fundamental aspects such as reduction and substitution, and we started to reason about confluence. As mentioned in \ref{subsec:beyond_conf}, there is still work to be done. There are two main research directions:
\begin{itemize}
    \item A relational analysis of \emph{termination}, already partly explored in \cite{backshouse-calculational-approach-to-mathematical-induction}, where a relational proof of of Newman's Lemma (which states that a terminating weakly/locally confluent reduction system is also confluent) is given. A possible future work is the study of termination of TRSs (e.g. recursive path ordering \cite{terese}).
    \item Together with the study of termination and confluence, another fundamental pillar of rewriting is the study of \emph{reduction strategies}. The sequential reduction we modeled in this manuscript is non-deterministic, while it would be useful (e.g. for applications on programming languages) to be able to model various reduction strategies. This can be done by further refining the derivative operator according to a certain strategy.
\end{itemize}

\bibliographystyle{plain}
\bibliography{main}

\end{document}